 \def\sdo{{\textit{SDO}}}
\documentclass[twocolumn,preprint]{aastex62}
\def\gsim{\ \raise 3pt \hbox{$\rangle$} \kern -8.5pt \raise -2pt \hbox{$\sim$}\ }
\usepackage{graphicx,natbib,color,amsmath,subfigure}
\usepackage{ulem}
\usepackage{pgfplots}
\newcommand{\blank}[1]{}
\usepackage{color}
\usepackage{mwe}
\newcommand{\gdf}{} %{\bf \color{blue}}
\newcommand{\gf} {} %{\bf \color{blue}}
\newcommand{\gfm}{} %{\color{blue}} %{\bf }
\def\mw{{microwave}}

\def\gs{{gyrosynchrotron}}

\begin{document}
%\maketitle

\title{Dynamics of the coronal magnetic field in the 2022-10-02 X-class flare}
%\titlerunning{Coronal magnetic field in an X-class solar flare}

%   \author{Gregory D. Fleishman\inst{1,2}
%             Tatyana Kaltman\inst{1}
%          Sijie Yu\inst{2}
%          }
% \institute{Institut f\"ur Sonnenphysik (KIS), Georges-Köhler-Allee 401 A, D-79110 Freiburg, Germany
% \and
% Center For Solar-Terrestrial Research, New Jersey Institute of Technology, Newark, NJ 07102, USA}

\author[0000-0001-5557-2100]{Gregory D. Fleishman}
	\affil{Center For Solar-Terrestrial Research, New Jersey Institute of Technology, Newark, NJ 07102, USA}

\affil{Institut f\"ur Sonnenphysik (KIS), Georges-Köhler-Allee 401 A, D-79110 Freiburg, Germany}

    \author[0000-0003-0366-1851]{Tatyana Kaltman }
    \affil{Institut f\"ur Sonnenphysik (KIS), Sch\"oneckstrasse 6, D-79104 Freiburg, Germany}

    \author[0000-0003-2872-2614]{Sijie Yu} %(\cntext{余思捷})}
	\affil{Center For Solar-Terrestrial Research, New Jersey Institute of Technology, Newark, NJ 07102, USA}

% 	\author[0000-0002-8078-0902]{Eduard P. Kontar}
% 	\affil{School of Physics \& Astronomy, University of Glasgow, G12 8QQ,
% Glasgow, United Kingdom}

%\begin{abstract}

   \date{today}

% \abstract{}{}{}{}{} 
% 5 {} token are mandatory
 
  %\abstract
  \begin{abstract}
  Solar flares are driven by release of free magnetic energy and often associated with restructurization of the magnetic field topology.  Yet, observations of evolving magnetic field in the flaring volume are limited to very few cases including the 2017-09-10 X8.2 limb flare; {\gfm thus, a verification of} whether a similar evolution takes place in other solar flares {\gfm is needed}. 
  {Here we report one more, {\gfm 2022-10-02, X1.1 class} solar flare but seen on disk, whose \mw\ data permit mapping the magnetic field over the flaring source and tracking magnetic field evolution over the course of the flare.
  We found that the coronal magnetic field shows a prominent decay with the rate up to 10\,G\,s$^{-1}$ in several (above) loop-top locations. The magnetic field is also confidently measured at the loop legs and the bottom part of the erupting filament. Prominent acceleration of electrons is detected where the magnetic field decays. We developed 3D models of the flare, whose magnetic field shows resemblance and also deviation from the magnetic field inferred from the \mw\ data.}
 {This study confirms that the coronal magnetic field decays during the rise phase of the solar flare. The amount of released magnetic energy is sufficient to support other components of the flare energy.}
\end{abstract}


  % \maketitle

%\keywords{Sun: Flares - Sun: X-rays, EUV, Radio emission}

\section{Introduction}
% {\gf blue: comments by GDF to be taken into account.}

% {\tk teal: text added by TK.}

% {\sy orange: comments by SY.}

% {\tmp red (tmp): either technical or preliminary text to be removed from the draft at a later stage. }

Solar flares {\gfm represent localized explosions in the solar atmosphere seen throughout the electromagnetic spectrum from}  a range of heights. Flares' transient brightenings may last from a few seconds to many hours and display a multitude of appearances over various phases of their development \citep{Fletcher2011, Benz2017}. 

In the framework of the ``standard'' solar flare model \citep[CSHKP,][]{Carmichael1964,Sturrock1966,Hirayama1974,Kopp1976}, an eruptive flux rope stretches the overlying magnetic field. This forms a quasi-vertical current sheet behind the flux rope. Magnetic reconnection and the shrinkage of field lines emerging from the reconnection region result in acceleration of a significant fraction of charged particles \citep{Krucker2010,Fl_etal_2011,Chen2020,2022Natur.606..674F}. These directly produce radio emission \citep{Bastian1998, 2011SSRv..159..225W} and, in denser regions, hard X-ray (HXR) and $\gamma$-ray emission \citep{Holman2011}. Some of the particles precipitate at the loop footpoints and generate electromagnetic radiation in various spectral ranges. This precipitation manifests itself vividly in the form of flare ribbons seen in the H$\alpha$ and UV ranges and in the form of bright HXR footpoints \citep{Hudson2011}. As a result of the impulsive plasma heating at the footpoints of the loops, chromospheric plasma evaporates and fills in the loops with hot plasma that shines in the soft X-ray and EUV ranges \citep{ Fisher1985}.  At the same time, the magnetic field lines outflowing from the reconnection site upward are ``wound'' on the curved eruptive flux rope, thereby decreasing the magnetic tension of the overlying loops and increasing the magnetic pressure at the bottom of the flux rope by additional poloidal flux facilitating its upward acceleration \citep{Forbes1995, 2000JGR...105.2375L, Liu2020}. 

Measuring plasma parameters, such as electron and ion density, temperature, velocity, and, most notably, the magnetic field during a solar flare at the sites of the magnetic energy dissipation is crucial for understanding and modeling the solar flare phenomenon \citep{2023ApJ...952..136A, 2023ApJ...946...46I}. 
However, exact locations, where, and the rates, at which, the magnetic reconnection takes place to drive the solar flare remain largely unknown despite decades of intense observational and theoretical study \citep{2000JGR...105.2375L, Priest2002, Su2013, 2017NatAs...1E..85W, Chen2020}. Only most recently, with the novel methodology of microwave imaging spectroscopy \citep{Gary_etal_2013, 2018ApJ...863...83G}, have the first ever detailed spatially resolved measurements of evolving magnetic field and directly accelerated nonthermal electron population been performed in a solar flare \citep{2020Sci...367..278F,Chen2020,2022Natur.606..674F}. These measurements confirm some facets of the standard picture of solar flares and reveal entirely new, apparently unexpected features.

Specifically, \citet{2020Sci...367..278F} reported the first dynamic measurement of the coronal magnetic field strength in a flare, which revealed a very fast decay of that field in the cusp region during its main phase. This decay is many orders of magnitude faster than the classical decay due to collisional Spitzer conductivity and, thus, requires strongly enhanced turbulent magnetic diffusivity \citep{FT_2013}. Another important finding reported by \citet{2020Sci...367..278F} is that the energy of nonthermal electrons accelerated in the flare increases synchronously in time with the magnetic field decay, which implies a direct link between the magnetic field decay (presumably, due to turbulent magnetic reconnection in an extended cusp region) and the process of particle acceleration.

Important details of this particle acceleration were reported by \citet{2022Natur.606..674F}.  They found that in the area where the magnetic field displays the fastest decay in time, literally all thermal electrons located there before the episode of the magnetic energy release (decay) are accelerated and converted to a nonthermal population in the volume.  This apparent 100\% efficiency of the acceleration process due to turbulent magnetic reconnection is striking, because models have not predicted such high acceleration efficiency. This points at a crucial role of the cusp region in the flare energy release and particle acceleration.

So far, very few solar flares were studied with this new technique. {\gf For example, \citet{2025ApJ...988..260F} studied a C-class flare with the same technique but did not find either prominent decay of the magnetic field or bulk acceleration of the electrons;} thus, it remains unclear how general the mentioned findings are, what is the typical magnetic field decay rate, what is the typical electron acceleration efficiency, how does the magnetic field inferred from the microwave imaging spectroscopy data compare with that in 3D models \citep[static or dynamic; e.g.,][]{Jiang2016, 2023ApJ...946...46I,Guo2024}, and many more.  The processes of energy conversion and particle acceleration cannot be conclusively modeled until a detailed physical picture of the acceleration site has been devised and tested through comparison with the appropriate observations. Although the results for the flare illustrated above were a key factor in testing a promising new model framework \citep{Arnold2021}, which couples equations for particle energy and transport with MHD equations, further progress will depend strongly on better observational constraints for the 3D flaring environment as applied to a large number of flares.

\begin{figure*}\centering
%\includegraphics[width=0.25%\textwidth]{aia_00032_211A_cr.png}
%\includegraphics[width=0.66\textwidth]{20221002_194000_680_230_aia00208.png}
% \includegraphics[width=0.3\linewidth]{AIA_EOVSA_1.png}
% \includegraphics[width=0.3\linewidth]{aia_00012_211.png}
% \includegraphics[width=0.3\linewidth]{aia_00000_94.png}
% %\includegraphics[width=0.758\linewidth]{AIA_EOVSA_3.png}
% \includegraphics[width=0.8\textwidth]{20221002_burst_time_withAIA.png}
%\includegraphics[width=0.95\textwidth]{fig-EOVSA-AIA.v2.export.pdf}
\includegraphics[width=0.95\textwidth]
{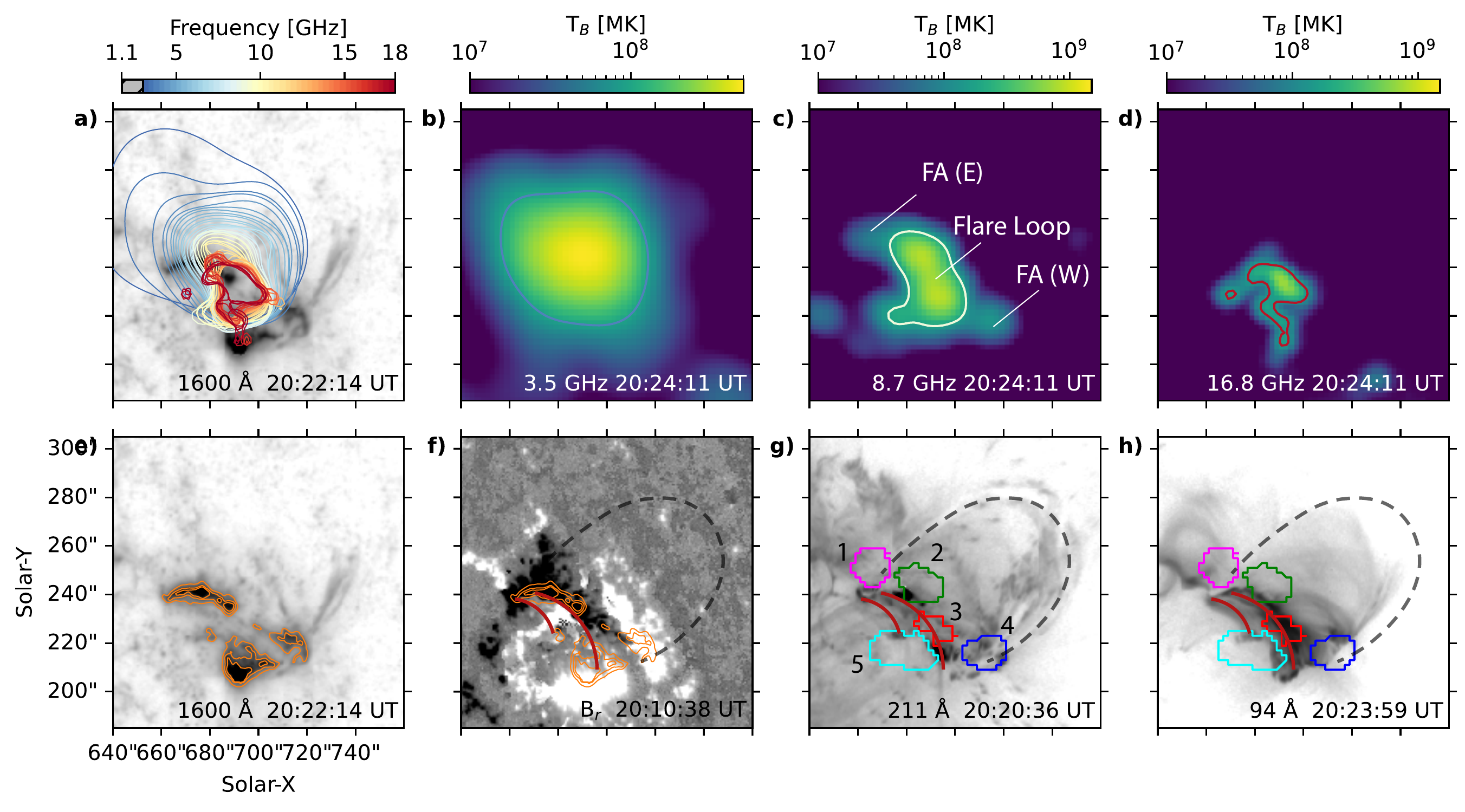}
\caption{Context of the flare location and associated filament eruption. 
(a) AIA 1600~\AA\ image with EOVSA multifrequency contours at 20\% of the maximum brightness at each frequency, color-coded by frequency from blue (low) to red (high) as indicated by the top color bar. 
(b--d) EOVSA images at 3.5, 8.7, and 16.8~GHz, each overlaid with the correspondingly colored contours from panel (a). 
(e) AIA 1600~\AA\ image with contours at 5, 13.6, 36.8, and 99.9\% of the maximum, highlighting the flare ribbons. 
(f) HMI LOS magnetic field with the same contours as in panel (e). 
(g--h) AIA 211~\AA\ and 94~\AA\ images with {\gf  colored and numbered} regions of interest (ROIs; see Section~\ref{S_bottom}) outlined. 
The dashed {\gf  black} curve marks the erupting filament, and the solid red lines indicate loops connecting the flare ribbons.}

\label{Fig:aia_eovsa}
\end{figure*}

\begin{figure*}\centering
%\includegraphics[width=0.25%\textwidth]{aia_00032_211A_cr.png}
%\includegraphics[width=0.66\textwidth]{20221002_194000_680_230_aia00208.png}
% \includegraphics[width=0.3\linewidth]{AIA_EOVSA_1.png}
% \includegraphics[width=0.3\linewidth]{aia_00012_211.png}
% \includegraphics[width=0.3\linewidth]{aia_00000_94.png}
% %\includegraphics[width=0.758\linewidth]{AIA_EOVSA_3.png}
 \includegraphics[width=0.8\textwidth]{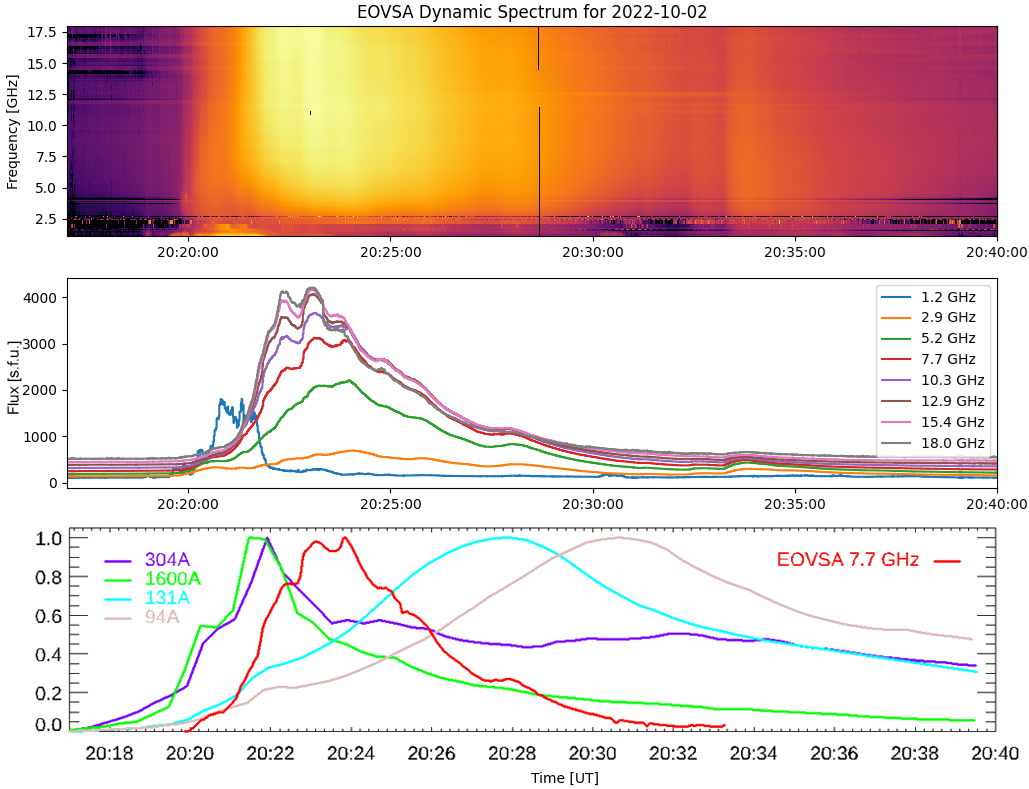}
\caption{Flare overview. 
First row: Dynamic microwave spectrum from EOVSA.
Second row: EOVSA total power light curves of the flux density (in sfu) at several frequencies.
Bottom: Normalized light curves at 304~\AA, 1600~\AA, 131~\AA, and 94~\AA \ (SDO/AIA) extracted from the region shown in Fig.\,\ref{Fig:aia_eovsa}, along with 7.7 GHz (EOVSA).
\label{Fig:EOVSA_overview}
}
\end{figure*}

% \begin{figure*}\centering
% \includegraphics[width=0.8\textwidth]%,bb=0 0 3000 1000,clip=0]
% {GF-20221002Xflare.jpg}
% \caption{
% \label{Fig:EOVSA_overview}
% }
% \end{figure*}

%\begin{figure*}\centering
%\includegraphics[width=0.98\linewidth]{profile_aia_eovsa_20221002_paper.png}
%\caption{
%{\tmp AIA+EOVSA  light curves}
%\label{Fig:light_curves}
%}
%\end{figure*}

%\subsection{Spectral fitting runs}

\section{Observations}
Here we investigate an X-class solar flare that occurred on 2022 October 2, beginning around 20:20 UT and lasting about 20 minutes in AR 13110 (centered at $x_c=695''$, $y_c=234''$). This event, associated with a filament eruption and a subsequent coronal mass ejection (CME), has been examined in several recent studies. For example, \citet{2023ApJ...952L...6S} showed that white-light kernels were co-spatial with microwave footpoint sources and co-temporal with hard X-ray emissions, suggesting a role for nonthermal electron beams. In a similar vein, \citet{2025ApJ...979L..43Y} demonstrated that energetic particles contribute directly to heating during this X1.0, white-light, flare.
Meanwhile,  \citet{2023ApJ...959...71D} reported on oscillations observed in a quiescent filament located at some distance, which were triggered by the EUV wave generated by the eruption. 

In this study, we focus on the radio observations of the flare, which exhibited a large microwave flux density exceeding 1000 solar flux units (sfu). A high flux density implies that the source {\gf  has either a high brightness temperature and/or} covers a larger area \citep[{\gf  e.g., Eq. 2 in}][]{2018ApJ...867...81F}; {\gf  in the latter case} allowing it to be sampled by a greater number of resolution elements — a crucial factor for robust model spectral fitting. The primary data come from the Expanded Owens Valley Solar Array (EOVSA; \cite{2018ApJ...863...83G}), complemented by context data from the \textit{Solar Dynamics Observatory}/Atmospheric Imaging Assembly (\sdo/AIA; \citet{2012SoPh..275...17L}) and the Helioseismic and Magnetic Imager (HMI) for 3D modeling.

\subsection{EOVSA}
\label{S_EOVSA}

The EOVSA  \citep{2018ApJ...863...83G}, a solar-dedicated radio instrument, provides
imaging spectroscopy observations at 451 frequencies between
1 and 18 GHz at a cadence of 1\,s. Raw data at 1\,s resolution were obtained from the public EOVSA data archive. Standard calibration procedures — including delay, bandpass, and complex gain calibration—were applied. Antenna gain corrections were subsequently refined using a self-calibration procedure (Cornwell and Fomalont 1999), following established methods for EOVSA. Self-calibration was performed using the compact radio source during the radio flux peak, and the derived solutions were applied to the entire 14-minute interval of interest. Final imaging was carried out at a 4 s cadence with 4 s integration across all frequency bands.

During the imaging reconstruction ({\tt CLEAN}) process, a circular restoring beam---the angular resolution of an image---was used. For frequencies up to 12 GHz, the full width at half maximum (FWHM) of the beam was set to 60$''f^{-1}$, where $f$ is the observing frequency in GHz; above 12 GHz, the FWHM was fixed at 5$''$,
{\gfm which was chosen to ensure adequate sampling of the synthesized beam given the fixed image pixel size of 2$''$ used at all frequencies. 
This choice guarantees at least $\sim$2.5 pixels across the FWHM of the beam even at the highest frequencies, which is important for stable image reconstruction. Using a smaller restoring beam would lead to undersampling and introduce systematic uncertainties without providing additional physically meaningful resolution.} 
In this manner, images were produced in 46 frequency bands, covering 2.5–17.8 GHz, for the period from 20:19 to 20:33 UT.

Total power calibration was then performed by adjusting the integrated flux in the image plane to match flare’s total power flux from independent single-dish measurements for each frequency band. {\gfm Each map was also quantified with a measurement uncertainty computed at 1$\sigma$ level as the standard deviation of the brightness distribution over the given brightness temperature map and converted to sfu using the pixel area $2''\times2''$.}

\subsection{SDO (AIA/HMI)}
{For the analysis of flare morphology and evolution, we used images with a pixel size of 0.6$''$ at UV (1600\,\AA) with a cadence of 24\,s and in EUV (94, 131, 211, 304, and 335\,\AA) wavelengths with a cadence of 12\,s from the AIA \citep{2012SoPh..275...17L}. The data were processed using the \texttt{aia\_prep.pro} routine in Solar Software (SSW). 
For NLFFF magnetic field reconstruction, we have used the HMI \citep{Scherrer2012} photospheric vector magnetograms  with a pixel size of 0.5$''$ and a cadence of 720\,s.
}

\section{Event Overview}
The X1.0 flare analyzed here is a classic two-ribbon event associated with a filament eruption, Figure\,\ref{Fig:aia_eovsa}, and strong microwave emission, Figure\,\ref{Fig:EOVSA_overview}. The eruption begins with the slow rise of the filament at 20:10 UT, followed by a rapid ascent at 20:20 UT, which marks the onset of the impulsive phase. This period is characterized by a pronounced increase in microwave emission (2.5–18 GHz) as electrons are accelerated by flare energy release. Downward-precipitating electrons heat the dense chromosphere, producing brightenings in (E)UV at or near the footpoints of the flare loops. Evaporation of chromospheric plasma further fills the coronal loops, yielding enhanced EUV emission.

{\gf
Figure \ref{Fig:aia_eovsa} presents contextual multi-wavelength observations of the event. The bottom row highlights the flare region: the flare ribbons, clearly visible in the 1600 {\AA} channel, are located on opposite magnetic polarities, as indicated by the photospheric line-of-sight (LOS)magnetic field. In the 211 and 94 {\AA} images, flare loops connecting the brightenings in the north and south ribbons are outlined by red solid lines, aiding visualization of the key structures discussed in the analysis. The erupting filament, connecting the two remote ends of the flare ribbons, is traced by a {\gf  black} dashed line. Additional brightenings associated with the filament eruption are visible southeast of the eastern filament anchor point (FA (E)). {\gf  Panel (g) displays five colored and numbered ROIs used in our analysis below.} The 94 {\AA} panel also illustrates the development of the post-flare arcade.

The top row shows EOVSA microwave contours at multiple frequencies (2.7––18 GHz) overlaid on an AIA\,1600 {\AA} image. At low frequencies, the source is extended and located cospatially with the north ribbon. At intermediate frequencies, a loop-like feature (white contour in panel c) connects the north and south flare ribbons, with the northern end also extending eastward toward FA (E). Two additional secondary sources appear east and west of the main flare arcade: the western source coincides with the western filament anchor point (FA (W)), while the eastern source is cospatial with the brightenings southeast of FA (E), though this feature is not discussed further in this study. At high frequencies, the emission becomes more compact, with the majority concentrated near the north ribbon and a smaller portion extending into the south ribbon.}

Figure~\ref{Fig:EOVSA_overview}  shows the temporal evolution of the flare, including the EOVSA dynamic spectrum (1–18 GHz), EOVSA flux density light curves, and AIA light curves at selected wavelengths, extracted from the region shown in Fig.\,\ref{Fig:aia_eovsa}. The 1600 and 304 {\AA} emissions rise rapidly around 20:20 UT, signaling the chromospheric response to precipitating ions and electrons. Microwave emission (2.5-18 GHz), presumably produced by nonthermal flare-accelerated electrons,  appears as the filament undergoes rapid liftoff at $\approx$20:20 UT. Around this time, a brief coherent radio burst is detected in the 1–2 GHz range, likely produced by plasma radiation from tens of keV electrons. Simultaneously, the 131 {\AA} emission increases, indicating the onset of plasma heating.

Both 1600 and 304 {\AA} emissions peak at approximately 20:22 UT, coinciding with pronounced brightenings along the flare ribbons. Meanwhile, the 131 and 94 {\AA} emissions, which trace hot post-flare loops at 10 and 6 MK respectively, show a delayed, gradual rise until about 20:31 UT before declining, reflecting the cooling of flare loops. The delay between the 94 and 131 {\AA} peaks further demonstrates this cooling process. The microwave emission reaches a broad maximum with multiple peaks between 20:22:30 and 20:24:00 UT, followed by a gradual decline.

The five regions of interest (ROIs) for microwave spectral analysis {\gf identified upon inspection of the inferred parameter maps, Sec.\,\ref{S_sp_fitting}, and} indicated by {\gf  numbered} colored contours in Figure~\ref{Fig:aia_eovsa}g, are as follows: {\gf  contours \# 1 (magenta) and 4 (blue)}  enclose the eastern {\gf  (FA (E))} and western {\gf  (FA (W))} footpoints of the erupting filament, {\gf  contour \# 5} (cyan) is co-spatial with the southern ribbon, and {\gf  ROIs \# 2} (green) and {\gf  3} (red) are located at/above the northern part of the loop---likely, ROI2 representing the ``above the loop-top'' (ALT) region and ROI3---at the loop-top region of the post-flare loops.

Overall, the evolution of microwave and EUV emissions aligns with the standard flare scenario: rapid particle acceleration and plasma heating during the impulsive phase, followed by gradual cooling. The spectral properties of the broadband microwave bursts in 2.5--18 GHz are consistent with gyrosynchrotron emission, enabling an in-depth spatially resolved spectral analysis of the evolving microwave source, as presented in the following sections. {\gf Only Fermi spectral X-ray data \citep{2023ApJ...952L...6S} but no imaging data are available, which, otherwise, would help constrain the flare model.}

%The EOVSA data file with overlapping 4~s time intervals and 2~s %cadence was produced for the model spectral fitting.
%(Goes, AIA, Eovsa) 

% \begin{figure*}\centering
% \includegraphics[width=0.98\linewidth]{profile_aia_eovsa_20221002}
% \caption{
% Normalized light curves from SDO/AIA (various colors as indicated in the legend, corresponding to the field of view shown in Fig.\,\ref{Fig:aia_eovsa}) and the EOVSA microwave light curve (black) at 12.61 GHz.
% \label{Fig:aia_211}
% }
% \end{figure*}

% \begin{figure*}\centering
% \includegraphics[width=0.98\linewidth]{aia_211_00032.png}
% \caption{
% AIA 211 Å filtergram overlaid with EOVSA contours (3–18 GHz, 20\% of the maximum level) and selected regions of interest, including the red (top of a loop) and gray (base of the erupted filament) areas.
% \label{Fig:aia_211}
% }
% \end{figure*}

% \begin{figure*}\centering
% \includegraphics[width=0.98\linewidth]{AIA_many.png}
% \caption{SDO/AIA images. {\tmp (i) add full caption; (ii) decide what figures to use and in what order.}
% \label{Fig:AIA_many}
% }
% \end{figure*}

% \begin{figure*}\centering
% \includegraphics[width=0.98\linewidth]{20221002_burst_time.png}
% \caption{EOVSA dynamic Spectrum  and light curves. {\tmp (i) add full caption; (ii) add more panels and make it entry overview Fig.?}
% \label{Fig:dynamic_spectrum}
% }
% \end{figure*}

% \begin{figure*}\centering
% \includegraphics[width=0.98\linewidth]{AIA_EOVSA.png}
% \caption{Left: SDO/AIA 335A . Right: SDO/AIA 1600A  with  EOVSA multispectral data.
% \label{Fig:AIA_EOVSA}
% }
% \end{figure*}

\section{Spectral model fitting}
\label{S_sp_fitting}

{\gdf The spectral model fitting is performed following the methodology described by \citet{Gary_etal_2013} using the GSFIT tool introduced in \citet{2020Sci...367..278F}. The results reported by \citet{Gary_etal_2013} are based on a large number of individual simulations and forward-fitting runs, while only one of which, the best one, was eventually described in the paper focusing on the ability of the methodology to truthfully recover the magnetic field in the flaring loops. As a result, unfortunately, several practically important settings and simulation outcomes remained undocumented in \citet{Gary_etal_2013}; we briefly summarize them here.  

The role and effect of finite frequency-dependent spatial resolution was investigated. The forward-fitting runs with different pixel sizes revealed that the physical parameters are recovered best of all for the smallest pixel size, 2\arcsec$\times$2\arcsec, roughly matching the array spatial resolution at highest frequencies. Degrading spatial resolution to match the spatial resolution at a lower frequency resulted in noticeably poorer parameter recovery due to progressively increasing role of the source nonuniformity.
To take into account and mitigate the effect of the frequency-dependent spatial resolution, which does not permit to resolve the source down to $2\arcsec\times2\arcsec$ scale at lower frequencies, frequency-dependent weights were introduced to give a progressively lower weight to lower frequency spectral data points. 
This is done by adding an additional factor, $A\times {\rm median}(\sigma(f))\times  f[{\rm GHz}]^{-1}$, to the observational errors $\sigma(f)$, where $A$ is a constant empirically selected as $A=4$. We adopt the same setting here.

\citet{Gary_etal_2013} employed the model spectral fitting with six free parameters, while reported only four of them; the outcome for the thermal number density, $n_{th}$, and low-energy cut-off, $E_{\min}$, remained undocumented, although they were recovered reasonably well in their simulations. The sensitivity of the \gs\ spectrum to these parameters is apparent from a supplementary video in \citet{2022Natur.606..674F} and employed recently to nail down an MeV-peaked electron component in a solar flare \citep{2026NatAs.tmp...10F}.
 }

% {\tmp Data files and fit files are stored in \\
% \verb|C:\Users\fleishman\Nextcloud\SM_Group\GS_FIT\2022-10-02|}

\begin{figure*}\centering
\centerline{\includegraphics[width=0.68\linewidth]{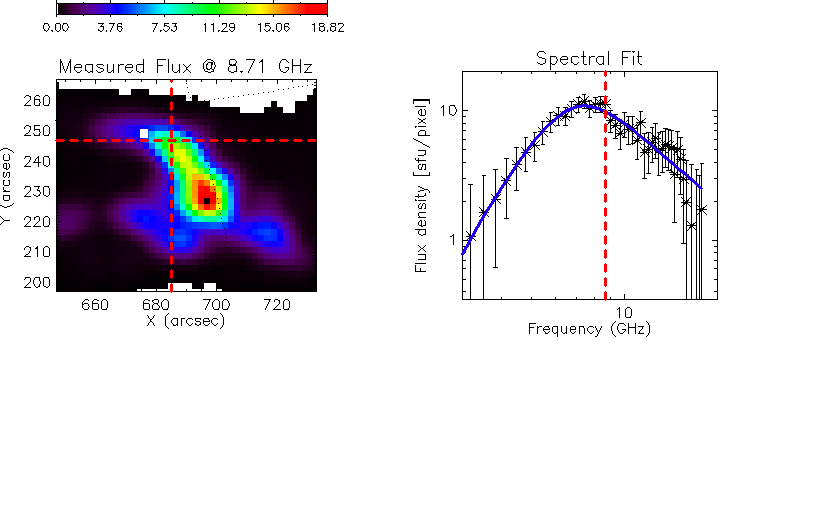}
\includegraphics[width=0.28\linewidth,bb=100 -120 430 120]{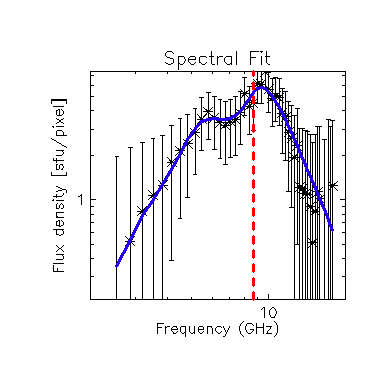}}
\caption{EOVSA image on 8.71 GHz 20:25:00\,UT (left) and observational spectra (symbols with error bars) from the pixel shown by red cursor (middle) in the left panel with coordinates $x=685\arcsec$, $y=247\arcsec$ {\gf and white pixel (right) with coordinates $x=675\arcsec$, $y=249\arcsec$} with the model spectral fits (blue lines). {\gf The statistical uncertainties of the measurements are {\gfm given at 1$\sigma$ level as explained in Section\,\ref{S_EOVSA}.}  An additional uncertainty {\gfm of $4\times {\rm median}(\sigma(f))\times  f[{\rm GHz}]^{-1}$,} clearly seen as increasing error bars towards lower frequencies, is added to the data to account for the frequency-dependent spatial resolution \citep{Gary_etal_2013}.} %The best fit parameters are: $B= 170\pm13$\,G, $\delta=3.9\pm0.1$, $n_{nth}=(7.3\pm1.9)10^8$\,cm$^{-3}$, $n_{th}=(1.1\pm7.9)10^9$\,cm$^{-3}$
\label{Fig:fitting}
}
\end{figure*}

{\gf
\subsection{Spectral fitting settings}
\label{S_fit_set}

Here we employ the spectral fitting tool, GSFIT \citep{2020Sci...367..278F}, which facilitates spectral model fitting of the spatially resolved \mw\ spectra. GSFIT requires two fixed settings{\gfm , the source depth $d$, and the minimum energy in the adopted single power-law (PWL) distribution of the nonthermal electrons $E_{\min}$. As most of the remote sensing methods, the \mw\ imaging spectroscopy offers few constraints that would help resolving the source along the LOS, so an assumption about the source depth is unavoidable. Its effect on the fit results is, however, minimal \citep[see results of modeling in][]{Gary_etal_2013} because the emission is only sampled along a limited portion of the LOS, where there are noticeable numbers of nonthermal electrons. The source depth is} here adopted at the default value $d=5.84$\,Mm that corresponds to the typical scale of 8\arcsec\ of the flare structures seen in Fig.\,\ref{Fig:aia_eovsa}. {\gfm  } 
The minimum energy {\gfm is adopted at} $E_{\min}=15$\,keV{\gfm , which is a typical value often reported from X-ray data analysis. Although it does affect the low-frequency spectral slope, \citep[see a supplemental video in][]{2022Natur.606..674F}, the extent and weights of the low-frequency spectra portion are insufficient to firmly constrain this parameter in our case}. We also fixed the maximum energy at $E_{\max}=5$\,MeV and plasma temperature at $T=20$\,MK {\gfm because our tests having them free showed that they are underconstrained by the data, provided that $E_{\max}\gtrsim1$\,MeV and $T\gtrsim5$\,MK}. Other parameters, the magnetic field $B$, its viewing angle $\theta$, the thermal number density $n_{th}$, the nonthermal number density $n_{nth}$, and the spectral index in the PWL distribution $\delta$ remained free to be determined from the model spectral fitting.}

\subsection{Spectral fitting runs}
\label{S_fitruns}

%{\tmp 2023-12-26. GF: I performed bulk spectral fitting of the EOVSA data file with overlapping 4~s time intervals and 2~s cadence.} 

{\gdf Although in a real source the observed emission is sampled along a LOS with varying physical parameters including the magnetic field,} the spatially resolved spectra are generally smooth and permit meaningful spectral fit in the entire spectral range 2.5-18\,GHz with a single GS source as illustrated by {\gf the middle panel of} Fig.\,\ref{Fig:fitting}{\gdf , which is an indication that the ranges of the parameter variation along a LOS are modest}. However, there are sub-areas, where many spectral instances show a double-peak structure{\gf ; see the right panel of Fig.\,\ref{Fig:fitting}} (see also Appendix and  Fig.\,\ref{Fig:Two_peak_fit} there). {\gdf Such spectral shape is an indication of two sources with distinctly different parameters rather than smoothly varying ones, which would result in a smoothly broadened spectrum. } To perform meaningful spectral fits of such cases we developed a new code  \citep[and compiled it to a new dll, which can be called by the spectral fitting tool GSFIT,][]{2020Sci...367..278F} that permits two independent uniform GS sources to be present within a given pixel. Therefore, we performed two main bulk spectral fitting runs---one with the standard single-source ``cost'' function and the other one with the new two-source function. {\gf For the latter case we adopted that the spectral indices of these two components are identical to each other and fixed $E_{\max}=1$\,MeV.} {\gfm This is done because now we have roughly two times smaller number of the constraining measurements per source; therefore, both $\delta$ and $E_{\max}$ are poorer constrained than for a single-source fitting. Inspection of the spectral fitting quality confirmed that the data do not require significantly different spectral indices for the assumed two sources.} We then applied an \textit{a posteriori} filtering as explained in Appendix to sort out the cases where the double source spectral structure with the low-frequency and the high-frequency sources is statistically significant from those, where only a single, main, low-frequency, source is present. We note that the spectral component with the higher peak frequency originates from sources with larger magnetic field and, thus, presumably located lower in the corona, compared with the low-frequency component. %From this perspective, we will refer to {\tk the second, ``low'' source, as a  high-frequency component, while the other  ``high'' one as a low-frequency (main) component}.     

%Such areas will be fit with either new two-source dll, or with appropriate restriction of the spectral range for the fitting.

%{\tmp The current run (still running) is for a selected ROI and full duration of about 15 minutes with variable flux thresholds: 20:19:52--20:20:04 Thr$=$1; 20:20:08--20:21:12 Thr$=$3; 20:21:16--20:27:52 Thr$=$20; 20:27:56--20:31:12 [213] Thr$=$7; 20:31:16 [214]--20:33:16 [234] Thr$=?$1; {\gf update thresholds!}

%I completed several fitting runs withe dlls mentioned above for 63 time frames that covers the rise, peak, and early decay phases of the flare. Analysis of the fit results, in particular, identification of cases when two-source fitting is needed, is under way. TK is developing an algorithm that analyzes the FIT spectrum and check if it has two significant spectral peaks or not. \textbf{TK: add description of the algorithm used.}}

\subsection{Spectral fitting results}

Once the fitting outcome has been classified between the single- and double- source cases we separated the data pertaining to the low-frequency (main) and high-frequency sources. Then, we complemented the main source data with those from the single source fitting. This means that our convention is that the single source data relate to the low-frequency source, which is justified by smooth transition of the parameters through boundaries of single and double source locations in the parameter maps. As a result, the parameter maps are continuous for the main, low-frequency source, while they have substantial gaps for the high-frequency source. 

\subsubsection{High-frequency source and regions of interest (ROIs)}
\label{S_bottom}

\begin{figure}\centering
\includegraphics[width=0.98\linewidth]{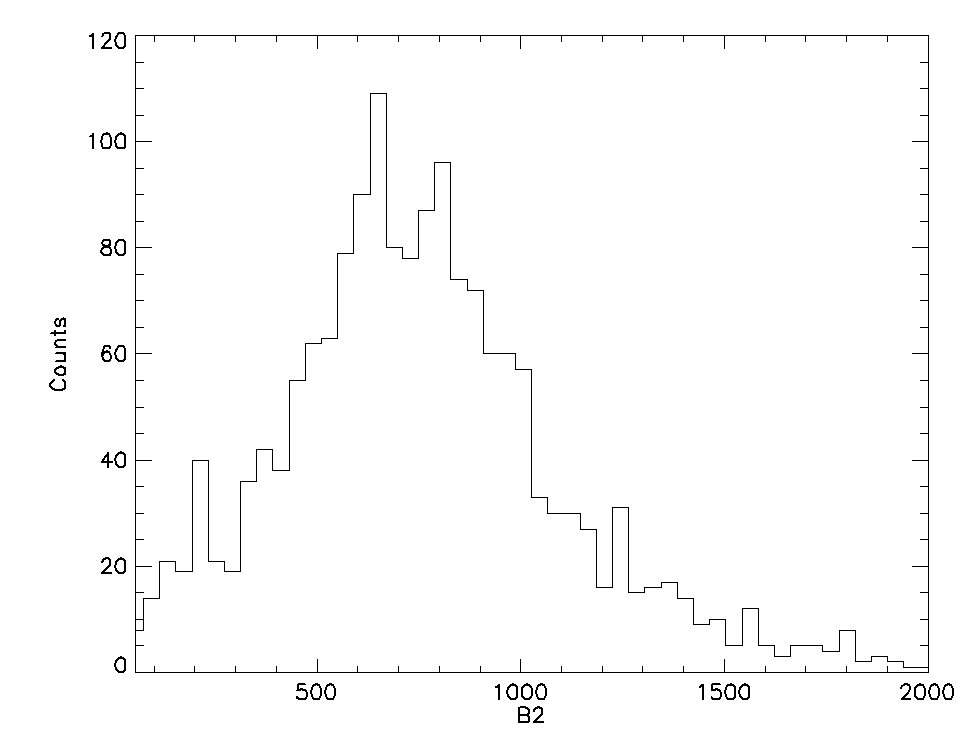}
\caption{Histogram of magnetic field values determined for high-frequency source over the entire analyzed duration of the flare. The field is predominantly between 400 and 1000\,G.
\label{Fig:B2_hist}
}
\end{figure}

\begin{figure}\centering
\includegraphics[width=0.98\linewidth]{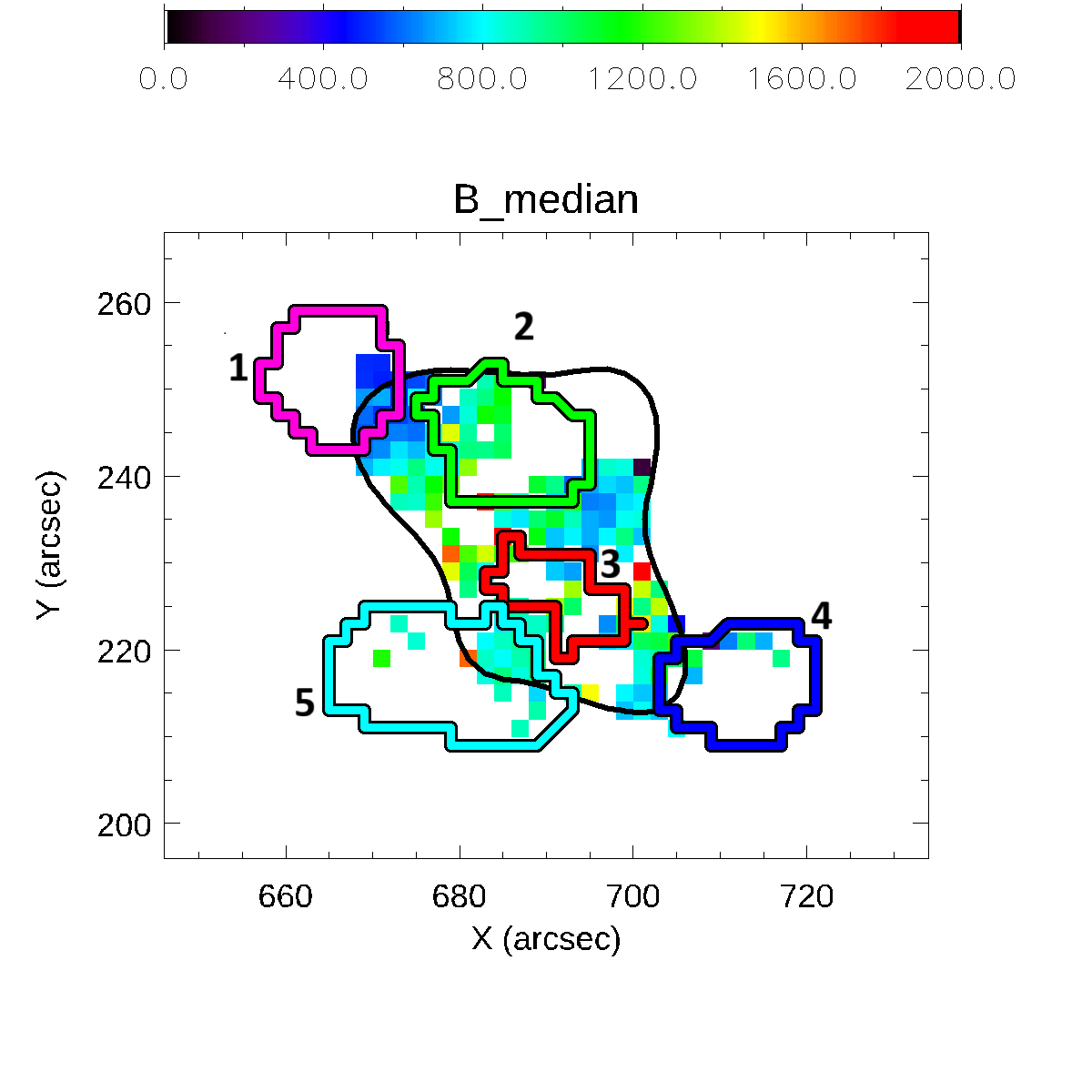}
\caption{Map of the regularized magnetic field values for the high-frequency source  from two-source fitting; see Section\,\ref{S_bottom} for details. Pixels where no two-source spectra was detected during the flare evolution are shown in white. The black contour represents EOVSA emission at 5.14 GHz at the 10\% level of the peak intensity at 20:20:16.0 UT, while the thick color contours indicate ROIs described in section \ref{S_bottom}.  
\label{Fig:B_mean_bottom}
}
\end{figure}

For the statistically significant cases of the double-source spectral fitting we created evolving maps of the physical parameters in the high-frequency source. Each individual map contains typically only a small number of pixels, where the second source is statistically confirmed. We did not identify any significant evidence of the magnetic field variation in time in this high-frequency source; the median value is roughly 700\,G; see Fig.\,\ref{Fig:B2_hist}.  For each pixel with the magnetic field values for the high-frequency source we determined the corresponding median values and formed a ``regularized'' map of the magnetic field in the high-frequency source, see Figure\,\ref{Fig:B_mean_bottom}. %{\gf Investigate cross-correlations between this map and the HMI maps / sample from the model.} --- No correlation found.

In this source, the thermal number density,  $n_{th}\sim10^{11}$\,cm$^{-3}$,  the nonthermal number density is relatively modest, $n_{nth}\sim10^{7}$\,cm$^{-3}$, while the spectral index evolves from $\delta\approx6$ to $\delta\approx4$ {\gfm during} the flare. These properties indicate that the high-frequency source is likely formed due to a leakage of a small portion of the nonthermal electrons down, to dense regions with relatively strong magnetic field from the main source(s) discussed below.

To investigate the parameter trends quantitatively, we selected five regions of interest (ROIs) indicated in Figure\,\ref{Fig:aia_eovsa}, which are also shown in Figure\,\ref{Fig:B_mean_bottom}. %Two of them (green and red) inscribe coronal / loop-top regions, two other inscribe the northern (magenta) and southern (cyan) loop legs, and one more (blue)--the footpoint / leg of the erupting filament, cf. Fig.\,\ref{Fig:aia_eovsa}. 
Note that our selected ROIs contain mainly ``white'' pixels as they are mainly dominated by single-source cases that do not contain the high-frequency source values.

\subsubsection{Coronal loop sources}

\begin{figure*}\centering
\includegraphics[width=0.98\linewidth]{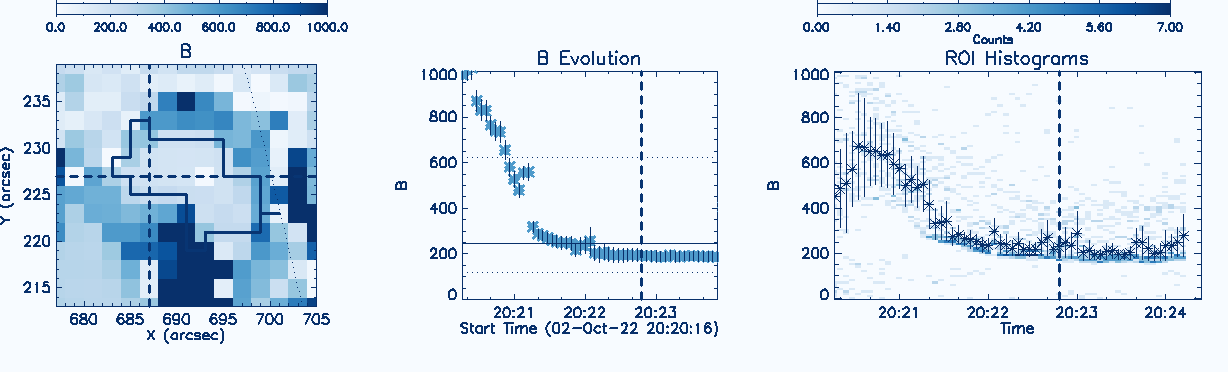}
%{20221002_202136-gsfit-Bmap-[21,15]Btimeplot-Broihist2d.png}
\includegraphics[width=0.98\linewidth]{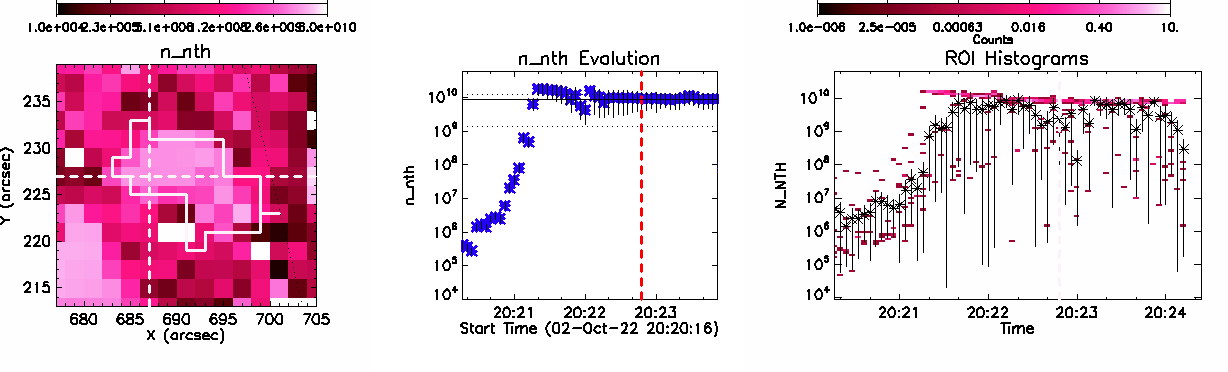}
%{20221002_202136-gsfit-N_NTHmap-[21,15]N_NTHtimeplot-N_NTHroihist2d.png}
\includegraphics[width=0.98\linewidth]{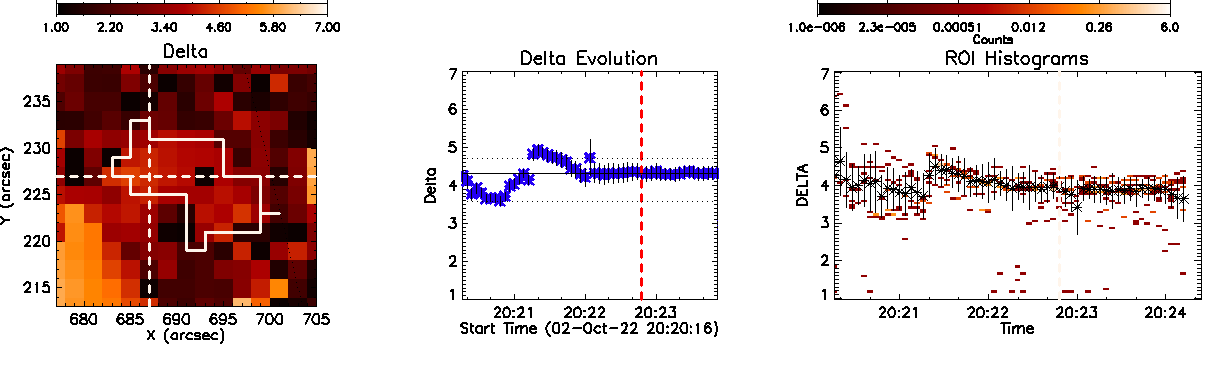}
%{20221002_202136-gsfit-DELTAmap-[21,15]DELTAtimeplot-DELTAroihist2d.png}
\caption{Left column: maps of the flare parameters inferred for 20:{\gf 22:48} UT zoomed-in on the southern loop-top ROI shown by the polygon: the coronal  magnetic field (top left), the nonthermal number density (middle left), and the spectral index $\delta$ (bottom left). Central column displays evolution of the corresponding parameter in the pixel shown by the cursor; {\gf vertical dashed lines show the time frame used to plot the parameter maps on the left.} The right column displays evolving 2D histogram of the parameters within the ROI {\gf shown by colored scattered points}. The black symbols with error bars show median values of the displayed parameters.  A rapid initial decay of the magnetic field followed by a slower decline is observed. The synchronized decay of the magnetic field and the rise in nonthermal electron density are evident. An animated version {\gfm of this Figure contains evolving} parameter maps {\gfm such as those shown in the left column of this Figure complemented by an evolving map of the thermal plasma number density. The length of the animation is 4 seconds.}
\label{Fig:gsfit_Bmap_Evo}
}
\end{figure*}

We identified two coronal locations, which are likely loop-top {\gf  (ROI3)} or above-loop-top ({\gf  ROI2}; more precisely, ``above the northern leg'') sources, associated with flux tubes highlighted in  Fig.\,\ref{Fig:aia_eovsa}. Both these coronal sources display prominent evolution of the magnetic field and parameters of the nonthermal electron population. Here, we focus on the loop-top source indicated by {\gf  ROI3}  (red). At the very beginning of the analyzed episode there is a trend of the magnetic field increase seen in the median values shown in the top right panel of Fig.\,\ref{Fig:gsfit_Bmap_Evo}, which is followed by a prominent decay of the magnetic field with the rate of $\dot{B}\approx10$\,G\,s$^{-1}$; see Fig.\,\ref{Fig:gsfit_Bmap_Evo}. This initial increase may be due to the magnetic field restructuring that triggers the flare and the filament eruption.   The forthcoming decay of the magnetic field with the rate of $\dot{B}\approx10$\,G\,s$^{-1}$  occurs synchronously with very fast increase of the nonthermal electron number density (see Fig.\,\ref{Fig:gsfit_Bmap_Evo}), which we interpret as prompt acceleration of the electrons due to dissipation of the free magnetic energy revealed by the decay of the magnetic field. {\gfm The thermal plasma number density is overall poorly constrained being $n_{th}\lesssim3\times10^{10}$\,cm$^{-3}$.} {\gf  It is easy to estimate that the rate of the decrease in the magnetic energy density is much higher than the rate of the gain in the nonthermal energy density during the flare rise phase, roughly 20:20:20---20:21:20 \,UT, which means that the decrease in magnetic energy is not balanced by the inferred increase in nonthermal energy similarly to the case reported by \cite{2020Sci...367..278F}; fig. 3B. This implies that other forms of energy, such as plasma motions and heating, turbulence, and waves, would play an important role in the energy budget; however, we do not have means to quantify them all.} 
%Accordingly, we conclude that the particle acceleration is not directly mediated by the release of magnetic energy, but rather by a chain of energy transformations that necessarily includes the inductive electric field and, perhaps, turbulence and other forms.}

It is interesting that the nonthermal electron number density reaches a peak value of about $(1-2)\times10^{10}$\,cm$^{-3}$, while the magnetic field continues to decay; however, the rate of this decay, $\dot{B}\approx2$\,G\,s$^{-1}$, significantly slows down at this stage. The decay ended in about a minute and then the magnetic field stays at roughly constant values, $B\sim200-400$\,G, specific for each location. {\gf  At this stage {\gfm it makes sense to perform a `sanity check' of sufficiency of the magnetic energy to confine the plasma in these post-flaring loops.} The magnetic energy density is now about $W_B\simeq1.5\times10^{3}$\,erg\,cm$^{-3}$, while the nonthermal energy density is about $W_{nth}\simeq0.5\times10^{3}$\,erg\,cm$^{-3}\approx0.3W_B$ {\gfm and a rough estimate of the upper bound of the thermal energy density for $n_{th}=3\times10^{10}$\,cm$^{-3}$ and $T=20$\,MK is $W_{th}\simeq0.25\times10^{3}$\,erg\,cm$^{-3}$}; thus, the plasma $\beta$ approaches{\gfm , but is smaller than} one in this stage \citep[cf., e.g.,][]{Krucker2010}.
} {\gfm We note that the nonthermal energy density is not well defined as its estimate relies on a postulated $E_{\min}$ value that are not well constrained by the data. The obtained relationship $W_{nth}\approx0.3W_B$ is consistent with the magnetic confinement and tells that $E_{\min}$ cannot be noticeable smaller than the adopted 15\,keV as the nonthermal energy scales as $E_{\min}^{1-\delta}$; larger $E_{\min}$ values are consistent with the magnetic confinement of the plasma. }

This process of the magnetic field decay and associated acceleration of electrons is not simultaneous at all points of these ROIs. In the ALT source (ROI2) the nonthermal electron number density reaches the peaks (at various pixels) between 20:21:12--20:21:20\,UT, while in the loop-top source (ROI3)--between 20:21:20--20:21:36\,UT. This implies that the flaring process originates at the ALT source and then propagates to the loop-top one. The thermal density cannot be reliably quantified in this case because its values have typically rather large uncertainties {\gfm or only an upper limit could be estimated as in several key regions of the 2017-09-10 flare \citep{2022Natur.606..674F}}; thus we cannot estimate the fraction of the electrons accelerated in this flare, although the measured values of the nonthermal electron number density at $(1-2)\times10^{10}$\,cm$^{-3}$ represent presumably a large fraction of the thermal electrons available in these ROIs before the flare{\gfm ; recall, the overall upper bound of the thermal plasma number density is $n_{th}=3\times10^{10}$\,cm$^{-3}$}.

The spectral index $\delta$ decreases and then increases during the phase of fast decay of the magnetic field (and increase of the nonthermal electron number density) revealing the soft-hard-soft spectral evolution often reported for the flare-accelerated electron population \citep[e.g.,][]{Holman2011}. At the next, microwave burst decay phase, the spectral index gradually decreases, indicating the spectrum hardening of the nonthermal electron population, which might be due to transport/trapping effects. This behavior takes place in both considered ROIs; the range of the spectral index variation is $\delta=3-5$ in both loop-top sources. %{\gf Add Fig./video.}{\tmp ---Do we have this Fig./video??}

% \begin{figure}\centering
% \includegraphics[width=0.98\linewidth]{20221002_202148-gsfit-Bmap-_20,15_Btimeplot.png}
% \includegraphics[width=0.98\linewidth]{20221002_202148-gsfit-N_NTHmap-[20,15]N_NTHtimeplot.png}
% \caption{{\gf Fix ROI color here.} {\kt Done.} Map of the coronal magnetic field (main/low-frequency source) inferred for 20:21:48 UT (top left), with a sample pixel marked by the intersection of red dashed lines inside the red loop-top ROI contour.  The light-blue contour represents EOVSA emission at 9.7 GHz at  the 10\% level of the peak intensity. Magnetic field evolution is shown in the top right panel. A rapid initial decay of the magnetic field followed by a slower decline is observed. A similar map for the nonthermal electron number density is shown (bottom left), along with its temporal evolution for the same pixel (bottom right). The synchronized decay of the magnetic field and the rise in nonthermal electron density are evident.
% \label{Fig:gsfit_Bmap_Evo}
% }
% \end{figure}

% \begin{figure}\centering
% \includegraphics[width=0.98\linewidth]{20221002_202148-gsfit-N_NTHmap-[20,15]N_NTHtimeplot.png}
% \caption{Map  of  the determined  non-thermal electron densities   for the loop top (up) ROI, an example its evolution for a pixel, and a Histograms. {\tmp (i) replace to white background; (ii) remove median data.} {\tk Done}
% \label{Fig:gsfit_NTh_Evo}
% }
% \end{figure}

\subsubsection{Southern leg/ribbon source}

Here (ROI5; cyan), in the leg source, many spatially resolved spectra have large uncertainties, thus, reducing confidence in the fit results. Some fits display a dichotomy of fit solutions resulting in a non-uniqueness of individual fit parameters. We will present general trends of the parameters without particular emphasis on their values.

The nonthermal electron number density is delayed here by 15--20\,s and increases with time more slowly, compared with those in the loop-top source (ROI3; red), but it reaches similar values of $(1-2)\times10^{10}$\,cm$^{-3}$ at the decay phase. Likely, this behavior is driven by the electron transport/escape from the loop-top source--a likely particle acceleration site. {\gf  The reported time delay is much longer than the time of flight; therefore, the transport has to be diffusive and mediated by turbulence/scattering on waves.} The spectral index $\delta$ decreases all the way with time from $\delta\approx9$ to $\delta\approx6$. %If this evolution is driven by the energy-dependent escape from the acceleration site, the high-energy electrons {\gf  has to be trapped longer than the} lower-energy electrons. {\gf  This may happen if high-energy electrons propagate in a strong diffusion regime mediated by turbulence, while the lower-energy electrons escape faster in a weak or moderate diffusion regime mediated by Coulomb collisions or their combination with scattering by turbulence/waves \citep[cf.][]{Fleishman2016}.} 
Overall, the median value of the magnetic field displays a rather modest decay from $\sim500$\,G to $\sim350$\,G over roughly 3 minutes; thus, $\dot{B}\lesssim0.8$\,G\,s$^{-1}$ with rather large uncertainties consistent with a constant magnetic field.

\subsubsection{Filament footpoints}

The remaining {\gf  ROI1 and ROI4} (magenta and blue)  inscribe the bottom parts of the erupting filament (including footpoints/anchors). Signal-to-noise ratio is not great in these ROIs and the fit solutions are not always unique; nevertheless, it permits obtaining rather smooth and overall consistent parameter maps. In the western footpoint {\gf  (ROI4)} the median value of the magnetic field slowly decreases with time from $\sim600$\,G to $\sim350$\,G. Here we cannot interpret this {\gf time variation} as a decay of the magnetic field {\gf in place} because prominent motions are present in this source; thus, a major role of the magnetic field advection is likely. The number density of the nonthermal electrons does not display any noticeable delay relative to the ALT source {\gf  (ROI2)}, although it is determined with large uncertainties; the median value is up to $\sim10^{10}$\,cm$^{-3}$. The spectral index $\delta$ varies mainly between 5 and 7. No reliable constraint can be derived for the thermal number density.

\begin{figure*} \centering
     \includegraphics[width=0.99\linewidth]{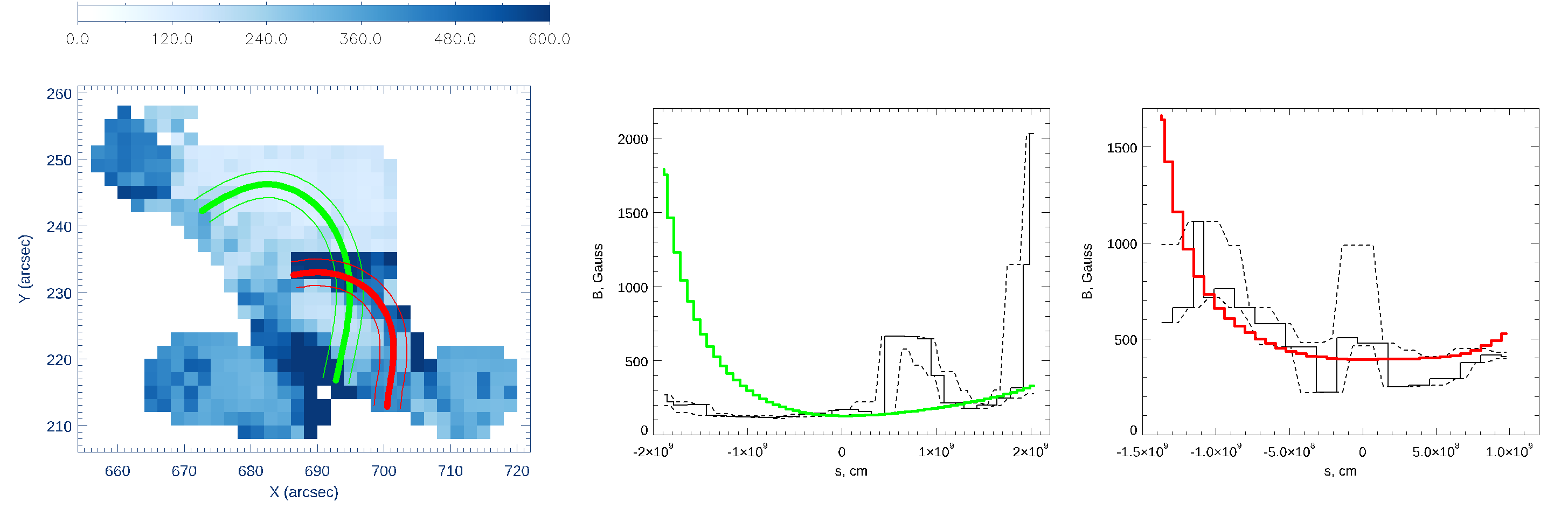} 
    \caption{%{\gf fix color table; perhaps, change to red temperature or so. Now both large and small values are black, which is misleading.} {\tk Done }
    Comparison of the regularized magnetic field map with the results of the 3D NLFFF reconstruction.    
    Left: The background shows the map of the regularized magnetic field values for the main source. Two reconstructed magnetic field lines are shown as thick red and green lines, each with a 1-pixel-wide tube around it, indicated by thin lines of the corresponding colors. 
    Right: the two plots correspond to the red and green reconstructed lines shown in the left image. Each plot compares the magnetic field values along the respective reconstructed line (red or green) with the values from the regularized map (solid black lines) and their uncertainties (dashed black lines). {\gf The starting points (negative $s$ coordinates) correspond to the upper-left footpoints of the field lines, while the end points are lower-right. Zero values corresponds to the minimal magnetic field along the line---a loop-top.}
    }
    \label{Fig:maglines_3}
\end{figure*}

In contrast, the eastern footpoint {\gf  (ROI1)} displays a roughly half-minute delay {\gf relative to the ALT source} in the nonthermal electron number density, but its values are rather modest, $n_{nth}\lesssim10^{7}$\,cm$^{-3}$. The spectral index is within $\delta=3-6$ and does not show any prominent evolution; thus, consistent with simple escape/precipitation of the electrons from the coronal sources. The magnetic field does not display here any statistically significant variation in time.
As far as we are aware, such {\gf diagnostics} have been obtained here for the first time for an eruptive filament.

\subsection{Regularized magnetic field map}
\label{S_regB}

The magnetic field was found to vary during the course of the flare, dominated by the field decrease in multiple flare locations. This decrease/decay is well pronounced during the flare rise phase, but it has finished at the flare decay phase; see Fig.\,\ref{Fig:gsfit_Bmap_Evo} as an example. Similarly to Section\,\ref{S_bottom}, here we obtain a median value of the magnetic field in each pixel at the stage where the magnetic field evolution is over to obtain a regularized magnetic field map as this action removes any outliers due to failed spectral fits. This regularized magnetic field map is displayed in the left panel of Fig.\,\ref{Fig:maglines_3}

\section{Modeling with GX Simulator}

To investigate the flare in the 3D domain, we employed the automated model production pipeline \citep[AMPP,][]{nita2023gx} to create three nonlinear force-free field extrapolations initiated by three \sdo/HMI photospheric vector magnetograms taken before the flare, during the flare peak, and after the flare. We found that all three represent a reasonable ground for the flare modeling; in particular, the ``peak'' model shows a reasonable resemblance with bright loops in the AIA images taken at the time of the flare peak. In what follows, we mainly employ the data cube created after the flare peak to make comparisons with the regularized magnetic field inferred from the \mw\ data at the flare decay phase, Sec.\,\ref{S_regB}. 

We employed two complementary approaches for making a comparison between the NLFFF model magnetic field values and those inferred from the spectral fitting. In the first approach, we computed several magnetic field lines in the 3D magnetic data cube crossing the \mw\ sources, projected these field lines onto the coronal magnetic field maps, and made direct point-to-point comparison as illustrated in Fig.\,\ref{Fig:maglines_3}. This comparison shows {\gf regions of both match and mismatch} between the magnetic field values inferred using two different methods. There are many possible reasons for such mismatches; for example, the \mw\ emission is formed over a range of heights along the LOS, where the nonthermal electrons are distributed, rather than at a single field line.

\begin{figure*}
    \centering
    \includegraphics[width=0.315\linewidth]{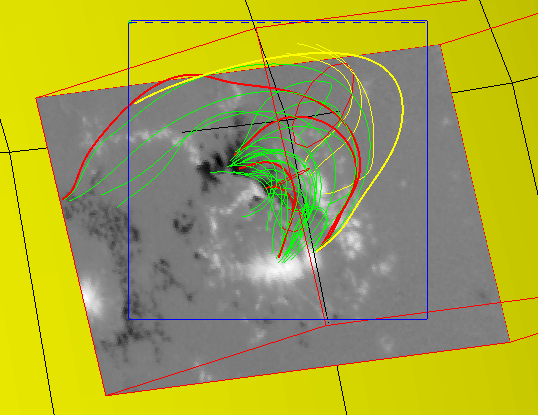}
    \includegraphics[width=0.3\linewidth]{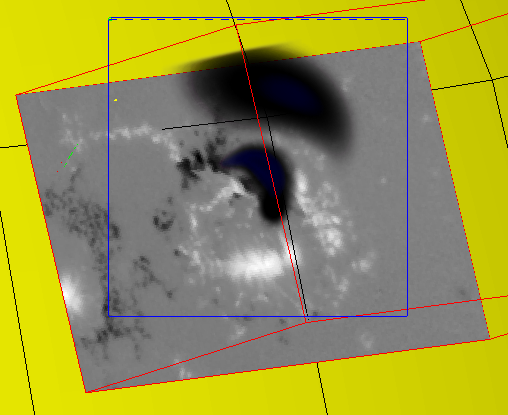}
    \includegraphics[width=0.31\linewidth]{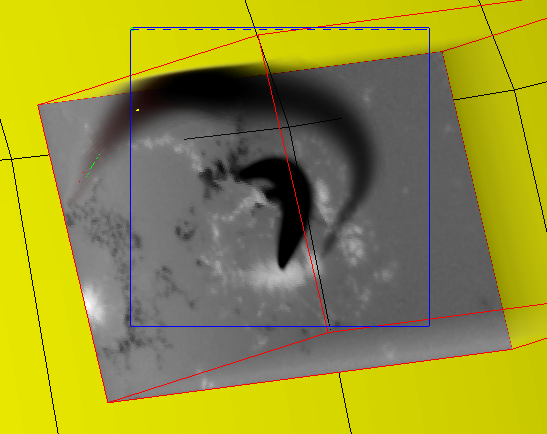}
    \includegraphics[width=0.31\linewidth]{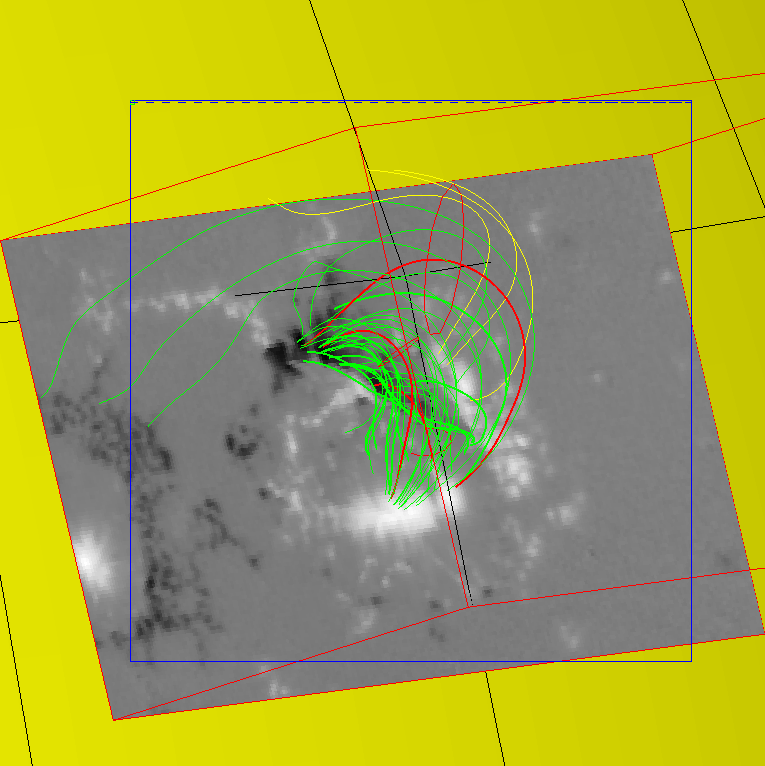}
    \includegraphics[width=0.31\linewidth]{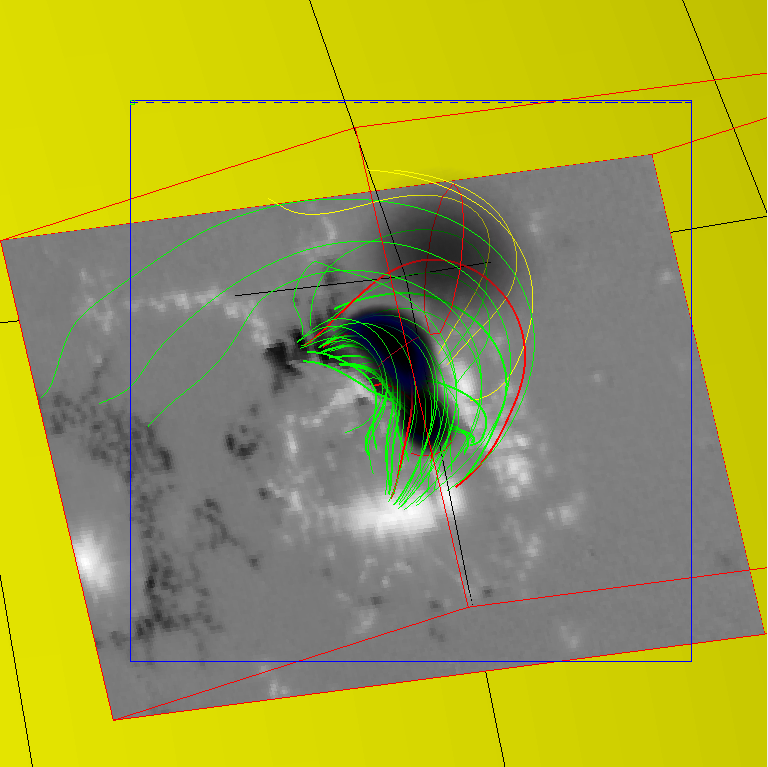}    \includegraphics[width=0.31\linewidth]{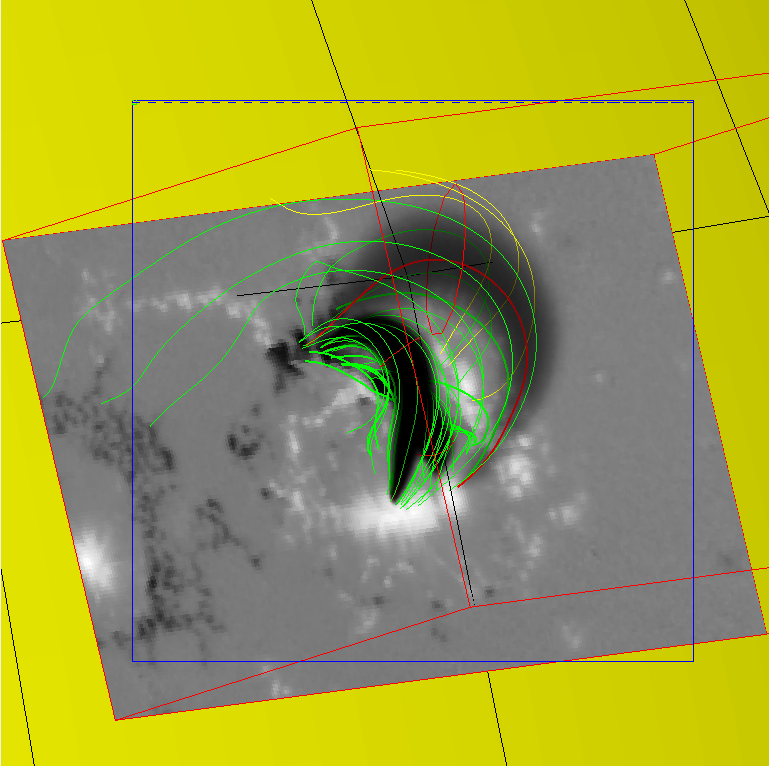}
    \caption{Visualization of two 3D models {\gf in the Earth view} for the 2022-10-02 solar flare, built within NLFFF data cubes generated at the peak of the flare (top row) and 12 minutes after the peak (bottom row), fine tuned to the microwave data at 20:22:20\,UT. Left to right: field lines/flux tubes; nonthermal electrons; and thermal electrons. In both cases at least three distinct flux tubes are needed.}
    \label{fig:model_3D}
\end{figure*}

\begin{figure*}
    \centering
    \includegraphics[width=0.31\linewidth]{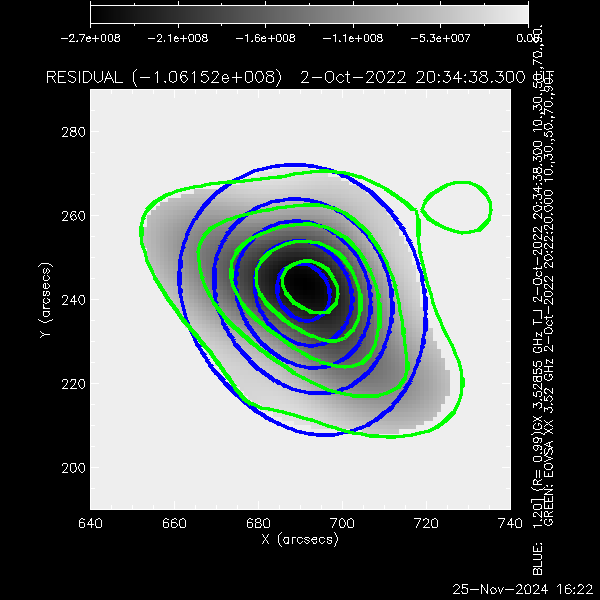}
    \includegraphics[width=0.31\linewidth]{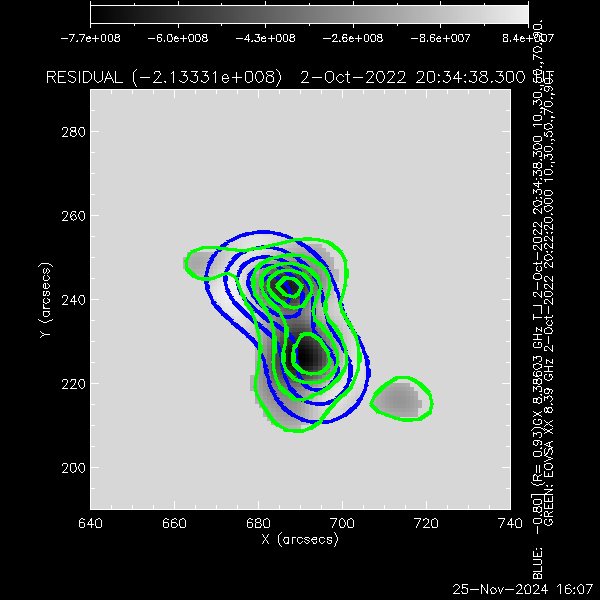}
    \includegraphics[width=0.31\linewidth]{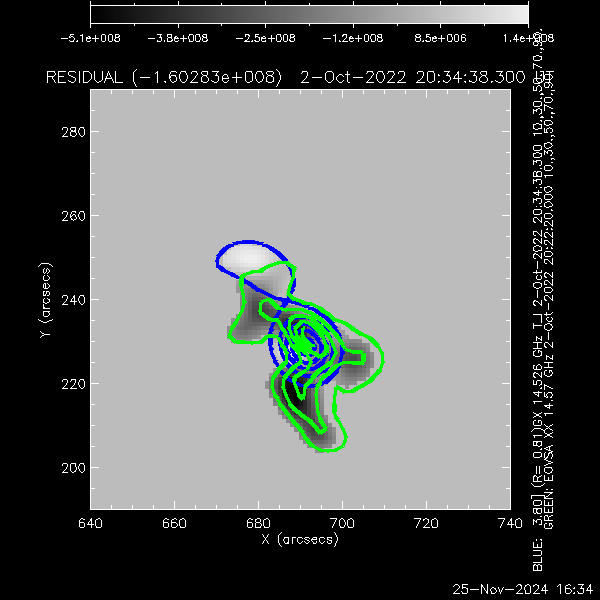}
        \caption{Model (blue contours) to data (green contours) comparison of the microwave images at three distinct frequencies 3.52 GHz, 8.39 GHz, and 14.5 GHz shown on top of the residual between these images. The cross-correlation coefficients are, respectively, $R=99,~93,~81\%$. }
    \label{fig:map_compar_after}
\end{figure*}

\begin{figure*}
    \centering
    \includegraphics[width=0.49\linewidth]{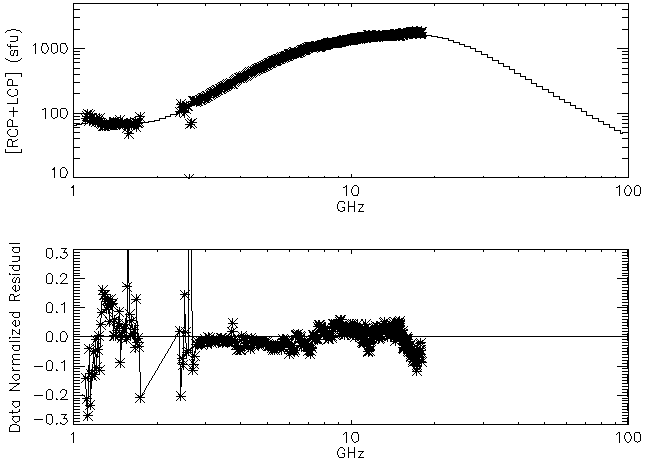}
    \includegraphics[width=0.49\linewidth]{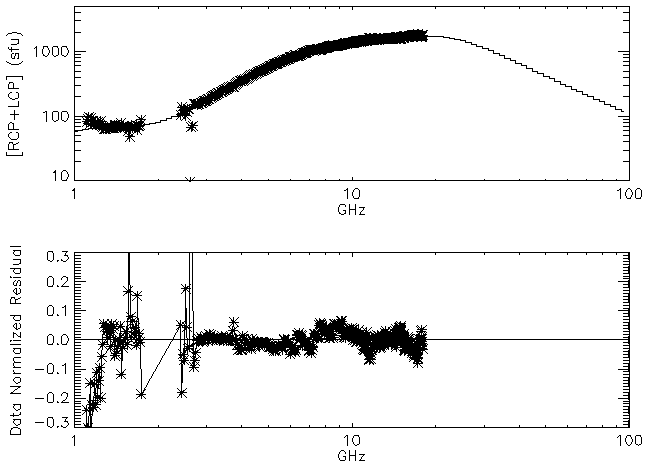}

        \caption{Model-to-data comparison of the spatially integrated spectra (top row) and their normalized residuals (bottom row). Left(Right): peak (after peak) time of the model data cube. }
    \label{fig:spectra}
\end{figure*}

To take this effect into account, we developed two 3D models of the flare, created using the NLFFF data cubes at the flare peak and decay phases; {\gf the model parameters are given in Table\,\,\ref{table_model_summary} and their visualization in Fig.\,\ref{fig:model_3D}}. Both models {\gf contain three distinct flux tubes similarly to the model developed in \citet{2025ApJ...988..260F}, two of which inscribe the field lines shown in Fig.\,\ref{Fig:maglines_3}. These models} 
reproduce both spectrum and multi-frequency images of the flare, while the quantitative metrics are marginally better in the case of the post-flare data cube.  {\gf  The numbers adopted in the model and reported in the Table enable a reader to closely reproduce the model if needed.} Figure\,\ref{fig:model_3D} displays the volume views of these two models, while Figure\,\ref{fig:map_compar_after} compares observed EOVSA images at several frequencies with the corresponding synthetic ones convolved with the EOVSA beam for the model built from the post-flare data cube. The spatial complexity of the source is best seen at third frequency due to higher spatial resolution there. Our simplified model cannot reproduce all fine details of the observed images; thus, calling for even larger number of the individual flaring flux tubes. Figure\,\ref{fig:spectra} shows a good match between synthetic and observed total power spectra  at the decay phase, 20:22:20\,UT, for both models, which validates the models even though there are some mismatches at lowest frequencies, especially in the ``peak'' model. {\gf We note that the models developed are not unique because of insufficient observational constraints. For example, having the X-ray imaging data complemented by the AIA data would help better constrain the model \citep{2021ApJ...913...97F,2023ApJ...953..174F}.
}
%using  GX models---one before,one after,  and one at the impulsive phase. So far, we made comparison of the regularized $B$ maps obtained for the decay flare phase, when the magnetic field evolution is over, with some field lines derived from ``after'' model. What is a potential use of two other models? Do we want to compare magnetic energies there? Or with the decrease of the magnetic energy obtained from the spectral fitting? In all three models NLFFF energy is smaller that PF energy. It is unclear if can make any solid conclusion from this.

%{\tmp I created 3-loop models ``before'' and ``peak,'' whose connectivities roughly match the EOVSA image at 8 GHz. ; some AIA channels are saturated. Once the *.ref file has been provided (Sijie), I will try to fine tune these model to match the data. If successful, we can compute weighted magnetic field map to be compared with the one derived from the model spectral fitting.}

%I created and fine tuned a model made from the ``peak'' NLFFF data cube, which very closely reproduces the total power spectrum at the decay phase, 20:22:20\,UT and also match images with the correlation coefficient up to 96\%, depending on frequency. The model is illustrated in Figure\,\ref{fig:model_3D}. Figure\,\ref{fig:map_compar_after} shows an example of the synthetic-to-observed image comparison at 9\,GHz with the cross-correlation coefficient $R=94\%$.

\begin{figure*}
    \centering
        \includegraphics[width=0.31\linewidth]{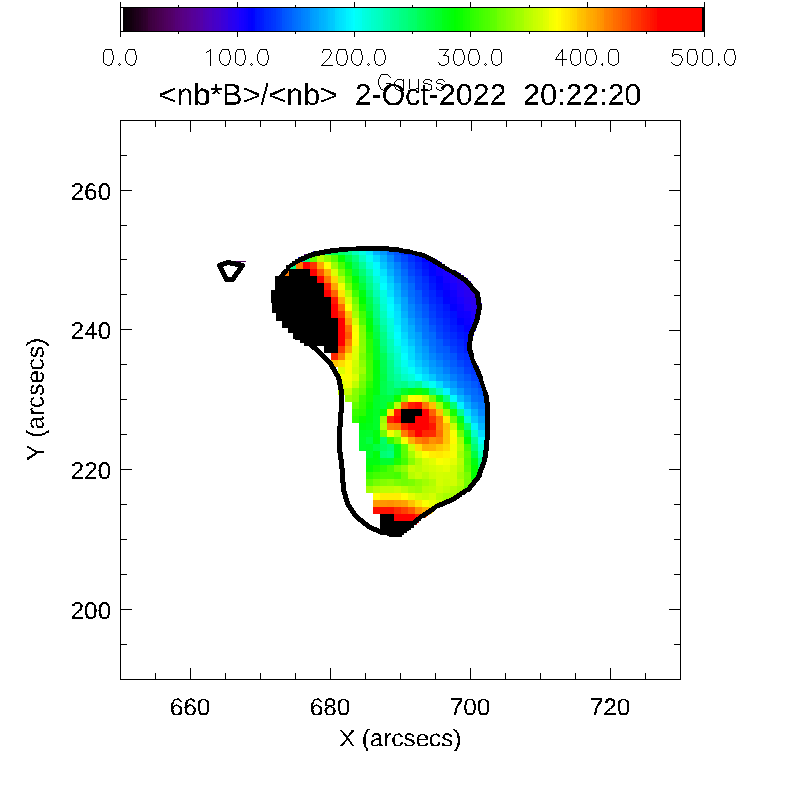}
    \includegraphics[width=0.31\linewidth]{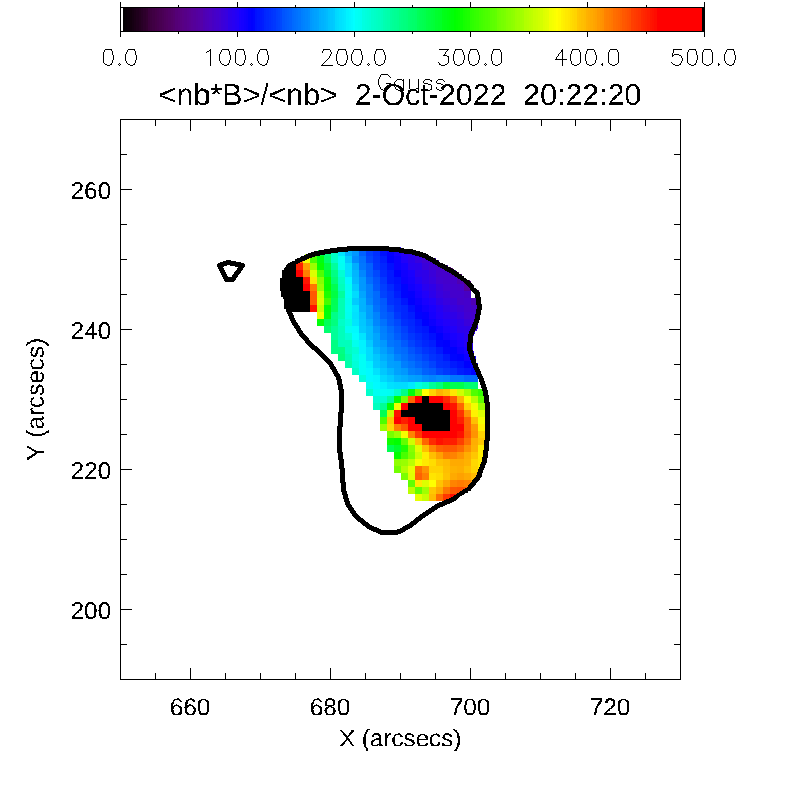}
    \includegraphics[width=0.31\linewidth]{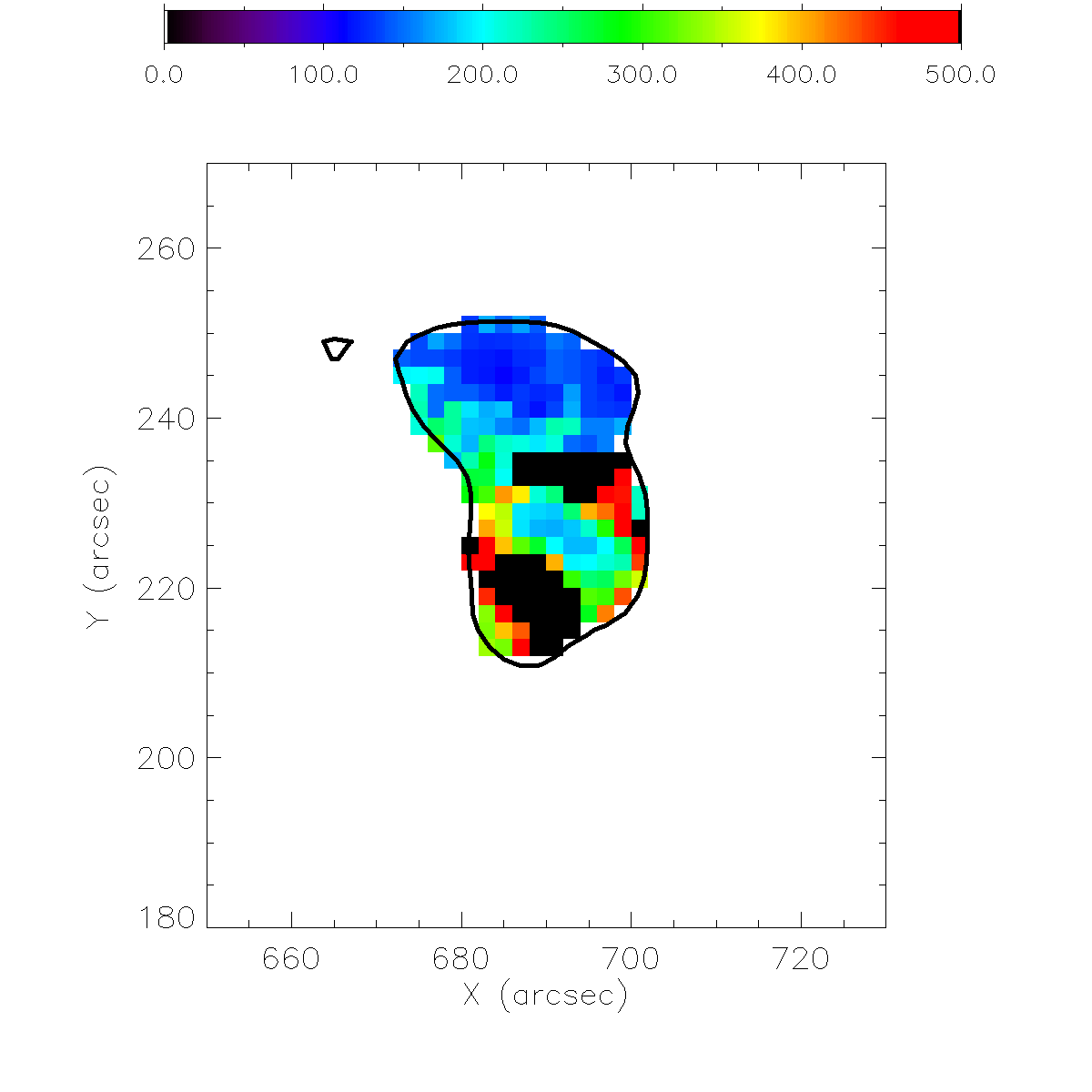}
        \caption{Magnetic field in the flaring loops weighted with the number density of the nonthermal electrons {\gf inside a region outlined by 15\% contour of the \mw\ brightness at 8.07\,GHz at 20:22:20\,UT}. Left (Middle): peak (after peak) time of the model data cube. Right: the regularized magnetic map obtained from the model spectral fitting of the \mw\ data. %The maps display similar, though not identical patterns.  %{\tmp (i) add full caption; (ii) Replace right panel to have white background and data within only EOVSA contour shown in the first panels (provide data to TK  {\tk Done}).  
        \label{fig:sampled_B}}
\end{figure*}

\begin{table*}[ht]
\caption{Parameters of the 3D model obtained from a post-flare NLFFF reconstruction. \hspace{1.7in}}
\begin{tabular}{l l l l l}
\hline\hline
Parameter & Symbol, units &  Loop 1 
 &  Loop       2
 &  Loop       3
\\ [0.5ex]
\hline
{\textit{Geometry}:} &  & \\
\quad Length of the Central Field Line     & $l$, Mm  &  50
 &  98
 &  22
\\
\quad Loop-top magnetic field & $B$, G & 128
 &  29
 & 354
\\
\quad Footpoint 1 magnetic field & $B$, G &1789
 &1827
 &1700
\\
\quad Footpoint 2 magnetic field & $B$, G &1779
 &1513
 & 931
\\
\quad Model Volume; $\left[\int n_0 dV\right]^2/\int n_0^2 dV$ & $V$, cm$^3$ &      1.31 $\times10^{27}$
 &      9.99 $\times10^{28}$
 &      6.72 $\times10^{26}$
\\
{\textit{Thermal Plasma}:} &  & \\
\quad Total Electron Number,  $\int n_0 dV$   & $N_0$ &      2.70 $\times10^{36}$
 &      6.64 $\times10^{36}$
 &      1.39 $\times10^{36}$
\\
\quad Emission Measure,  $\int n_0^2 dV$   & $EM$, cm$^{-3}$ &      5.59 $\times10^{45}$
 &      4.42 $\times10^{44}$
 &      2.87 $\times10^{45}$
\\
\quad Mean Number Density,  $EM/N_0$   & $n_{\rm th}$, cm$^{-3}$ &      2.07 $\times10^{ 9}$
&        6.65 $\times10^{ 7}$
&        2.07 $\times10^{ 9}$
\\
\quad Temperature   & $T$, MK &20
& 5
&20
\\
\quad Thermal Energy   &  $W_{\rm th}$, erg &      2.24 $\times10^{28}$
 &      1.37 $\times10^{28}$
 &      1.15 $\times10^{28}$
\\
{\textit{Nonthermal Electrons}:} &  & \\
\quad Total Electron Number,  $\int n_{\rm b} dV$   & $N_{\rm b}$ &      2.05 $\times10^{36}$
 &      1.23 $\times10^{34}$
 &      6.81 $\times10^{35}$
\\
\quad Mean Number Density,  $\int n_{\rm b}^2 dV/N_{\rm b}$   & $n_{\rm nth}$, cm$^{-3}$ &      5.55 $\times10^{ 9}$
 &      2.20 $\times10^{ 6}$
 &      5.24 $\times10^{ 9}$
\\
\quad PWL Energy Range & $E$, MeV &     0.015 $-$ 10.
 &     0.020 $-$ 0.62
 &     0.015 $-$ 10.
\\
\quad PWL Spectral Index & $\delta$ &      3.99
 &      3.00
 &      3.87
\\
\quad Nonthermal Energy    & $W_{\rm nth}$, erg &      7.40 $\times10^{28}$
 &      7.90 $\times10^{26}$
 &      2.51 $\times10^{28}$
\\
[1ex]
\hline
\end{tabular}
\footnote{ This model matches the \mw\ spectrum and images. %; see Table\,\ref{table_model_summary}. %The Table is made in the IDL code\\ \verb|d:\GX_MODELS\2022-10-02\Table4Latex\table4latex.pro|.
}
\label{table_model_summary}
\end{table*}

These validated models now can be used to compare the magnetic field values inferred from the \mw\ model spectral fitting with those in our 3D models. To this end, we have to compute the maps of the magnetic field $B_{weighted}(x,y)$ in the model weighted with the spatial distribution of nonthermal electrons $n_{nth}(x,y,z)$:
\begin{equation}
B_{weighted}(x,y) = \frac{\int\limits_0^{h_{\max}}B(x,y,z)n_{nth}(x,y,z)dz}{\int\limits_0^{h_{\max}}n_{nth}(x,y,z)dz}, 
\end{equation}
and then compare these maps with the regularized magnetic map obtained for the decay flare phase. This weighted magnetic field is relevant because the \mw\ emission is mainly produced in the volume filled with the emitting nonthermal electrons. Figure\,\ref{fig:sampled_B} shows two sampled maps, {\gf with the contribution from the longest loop (loop 2 in the table) excluded,} obtained from two our models and the regularized map of the inferred magnetic field. {\gf The contribution from loop 2 was excluded because it affects only the lowest frequencies, while the spectral fitting results are mainly sensitive to higher-frequency data \citep[cf.][]{2025ApJ...988..260F}.} All maps show similar ranges of the magnetic field and similar trends of the magnetic field increase from the top of the maps to the bottom, while details of the maps are different. These differences are, likely, due to both inaccuracies and non-uniqueness of our idealized 3D models and uncertainties of the model spectral fitting of the \mw\ data.

\section{Discussion}

We have analyzed \mw\ imaging spectroscopy data of the X-class 2022-Oct-02 solar flare seen on disk. The model spectral fitting of these \mw\ data reveals magnetic energy release manifested by a rather fast decay of the coronal magnetic field with the rate up to 10\,G\,s$^{-1}$ in the loop-top or above-loop-top regions at the flare rise and peak phases. This decay is temporally correlated with the increase of the number density of the nonthermal electrons accelerated, presumably, due to the release of the free magnetic energy. A significant portion of available electrons gets accelerated. These findings are consistent with those for another, 2017-Sep-10, X-class solar flare studied earlier with the same methodology \citep{2020Sci...367..278F, 2022Natur.606..674F}{\gf , although a weaker C-class flare studied in \citet{2025ApJ...988..260F} is different. It does not display either prominent decay of the magnetic field or bulk acceleration of electrons, which calls for a systematic study of the dependence of these trends on the flare class.}

Unlike the 2017-Sep-10 solar flare that occurred on limb, the 2022-Oct-02 flare was seen on disk; thus, permitting 3D models initiated by the photospheric vector magnetograms to be created. This enables direct comparison between the magnetic field inferred from observations and obtained in the modeling. Our tests demonstrate that the two magnetic fields are in a reasonable agreement between each other.

With the inferred {\gf flare parameters} and the developed model, we are in the position to estimate the energy components---the released free magnetic energy and the energy of the accelerated nonthermal electrons. The released magnetic energy, $W_B$, can be estimated as the difference between the peak magnetic energy before the magnetic field started to decay and the final steady-state magnetic energy integrated over the source volume. Making this exercise for the two loop-top sources, we obtain $W_B\sim 3\times10^{31}$ and $\sim1.5\times10^{31}$\,erg for the ALT and loop-top sources, respectively. The %corresponding instantaneous 
nonthermal energy integrated over these two sources is about $\sim6\times10^{29}$\,erg at the flare peak time. This is consistent with the nonthermal energy obtained with the integration of the model flux tube volumes, $\sim10^{29}$\,erg, devised for an early decay phase, at 20:22:20\,UT, see Table\,\ref{table_model_summary}.\\

%What else?

%Should we discuss measurement at the filament?

%I sampled the model magnetic field with the spatial distribution of $n_{nth}$; this yields the magnetic field roughly a factor of 2 larger than the one derived from the model spectral fitting of EOVSA data. It is interesting to create a model from the ``after'' 3D data cube.

%I have created a new 3D model from the ``after'' NLFFF extrapolation. The spectral match in this case is even better; image match is marginally better. I have yet to sample the model volume to make comparison with Fig. 8.

%If a change of morphology is found, this would imply change of magnetic connectivity due to restructurization of the magnetic field. Need to be checked by the analysis of sequence of 45-s Blos maps.

%\acknowledgments

EOVSA was designed, built, and is now operated by the New Jersey Institute of Technology (NJIT) as a community facility. EOVSA operations are supported by NSF grant AGS-2436999 to NJIT.
This work was supported in part by NSF grant 
AGS-2425102,  %New Flare NSF Grant
and NASA grant
80NSSC23K0090 %(Dale's new HSR grant)
to New Jersey Institute of Technology. TK was supported by DFG grant eBer-24-58553.

%\bibliographystyle{alpha}
%\bibliography{fleishman,BIB_20221002,references,dop_refer}
%\bibliographystyle{aa} % style aa.bst
%\bibpunct{(}{)}{;}{a}{}{,} % to follow the A&A style

%\appendix

\section*{APPENDIX: Filtration of Various Spectral Cases %\sout{Classification of Spectra by Number of Peaks}
}

\begin{figure}\centering
\includegraphics[width=0.99\linewidth]{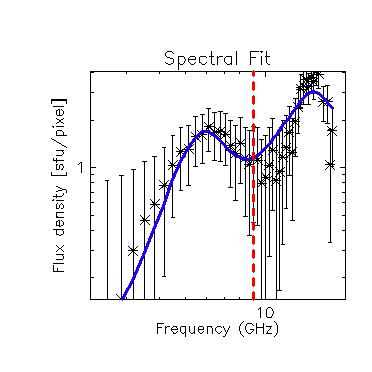}
\caption{Two-peak fit example (blue line): a spectrum (symbols with error bars) with two peaks from a single pixel suggests the statistically significant  presence of two sources along the line of sight.
%{\tmp (i) replace to a white background; (ii) move to appendix?} {\tk Done}
\label{Fig:Two_peak_fit}
}
\end{figure}

As has been explained in Section\,\ref{S_fitruns}, there are cases that cannot be described by a single-source spectral model such as shown in Fig.\,\ref{Fig:fitting}, while require two distinct sources as in Fig.\,\ref{Fig:Two_peak_fit}. It is difficult to distinguish those cases algorithmically at the run time of the model spectral fitting. For this reason, we performed two independent spectral model fitting runs, one with a single and the other with two sources, over the entire area and the time range of interest, and then filtered out different cases \textit{a posteriori} as described below.

At the first step, for each case (i.e., for each pixel at each time frame) we counted the number of extrema---minima and maxima of the spectral model (like that shown by blue lines in Figs.\,\ref{Fig:fitting} and \ref{Fig:Two_peak_fit}). Apparently, this procedure identified spectra with no extreme (excluding the starting and ending spectral points), or one, two, or three extrema. 

At the second step, we designated all cases with one or no spectral peak to represent a single-source and marked up the corresponding case accordingly. We separated the cases with two extremes and three extremes and treated them differently. 

At the third step, we analyzed the spectra with two extremes. These spectra display a spectral peak at intermediate frequencies and a rising branch at high frequencies. 
{We verified whether the difference between the maximum and minimum values in the fitted or observed spectrum exceeded both measurement errors at the maximum and minimum points. If so, the spectrum was classified as having a confirmed maximum-minimum pair (a two-source case); otherwise, it was marked as having a single maximum (a single-source case). }

At the fourth step we analyzed cases with two spectral peaks to figure out if the presence of two spectral fits is statistically significant or not.  To do this, {we checked the difference between the smaller of the two maxima and the minimum. If this difference (in the observed or fitted spectra) exceeded both measurement errors at the maximum and minimum points, the spectrum was classified as belonging to a two-source case.}

At the final step, after filtering the cases out, we combined the fit results in a single database. {When the described criterion confirmed the presence of the second (high-frequency) source in the spectrum, we stored the parameters from the two-source spectral model fit for both the low-frequency and high-frequency sources. In all other cases, we used the parameter values of the single (low-frequency) source obtained from the single-source spectral model fit.}

\end{document}